\documentclass[a4paper,11pt]{article}
\pdfoutput=1 
\usepackage{jheppub} 
\usepackage{shuffle}
\usepackage[T1]{fontenc} 
\usepackage{lmodern}

\usepackage{float}
\usepackage{multirow}
\usepackage{bbm}
\usepackage[table]{xcolor}
\usepackage{mathtools}
\usepackage{mathdots}
\usepackage{amsmath}
\usepackage{bm}
\usepackage{amsfonts}
\usepackage{dsfont}
\usepackage{amsthm}
\usepackage{enumitem}
\usepackage{booktabs}
\usepackage{siunitx}
\usepackage{arydshln}
\usepackage{hyperref}

\usepackage{graphicx}

 \usepackage{tikz}
    \usetikzlibrary{positioning}
    \usetikzlibrary{arrows}
    \usetikzlibrary{shapes}
    \usepackage{pgfplots}
    \usetikzlibrary{calc}
    \usetikzlibrary{decorations.markings}
     \usetikzlibrary{decorations.pathmorphing}
    \tikzset{snake it/.style={decorate, decoration=snake}}
\def\centerarc[#1](#2)(#3:#4:#5) % Syntax: [draw options] (center) (initial angle:final angle:radius)
    { \draw[#1] ($(#2)+({#5*cos(#3)},{#5*sin(#3)})$) arc (#3:#4:#5); }
    \usepackage{booktabs}

\newcommand{\tr}{{\rm tr}}

\definecolor{cnblau}{RGB}{0,80,147}

\newcolumntype{?}{!{\vrule width 1.5pt}}
\newenvironment{gleichung*}{\begin{equation*}\begin{aligned}}{\end{aligned}\end{equation*}}

\newcommand{\eps}[0]{\epsilon}

\def\beq{\begin{equation}}
\def\eeq{\end{equation}}
\def\bsp#1\esp{\begin{split}#1\end{split}}

\usepackage[xindy,toc]{glossaries} 

\usepackage{subcaption}

\usepackage[perpage]{manyfoot}
\usepackage{mwe}
\usepackage{braket}
\usepackage{slashed}
\usepackage{makecell}
\usepackage{listings}
\newcommand{\tth}{$gg \to t \overline{t} H$ }
\newcommand{\GIT}{\href{https://github.com/p-a-kreer/TTH}{git clone https://github.com/p-a-kreer/TTH.git}}

\title{One loop QCD corrections to $gg \to t\overline{t}H$ at $\mathcal{O}(\epsilon^2)$}

\author[a]{Federico Buccioni,}
\author[a]{Philipp Alexander Kreer,}
\author[b]{Xiao Liu,}
\author[a]{Lorenzo Tancredi}

\affiliation[a]{Physics Department, Technical University of Munich, D-85748 Garching, Germany}
\affiliation[b]{Rudolf Peierls Centre for Theoretical Physics, Parks Road, Oxford OX1 3PU, UK}

\emailAdd{federico.buccioni@tum.de}
\emailAdd{philipp.a.kreer@tum.de}
\emailAdd{xiao.liu@physics.ox.ac.uk} 
\emailAdd{lorenzo.tancredi@tum.de} 

\abstract{
We compute the one-loop corrections to \tth up to order $\mathcal{O}(\epsilon^2)$ in the dimensional regularization parameter.  
We apply the projector method to compute polarized amplitudes, which generalize
massless helicity amplitudes to the massive case. 
We employ a semi-numerical strategy to evaluate the scattering amplitudes. 
We  express the form factors through scalar integrals analytically, 
and obtain separately integration by parts reduction identities in compact form. 
We integrate numerically the corresponding master integrals with an enhanced implementation of the Auxiliary Mass Flow algorithm.  
Using a numerical fit method, we concatenate the analytic and the numeric results, 
to obtain fast and reliable evaluation of the scattering amplitude. 
This approach improves numerical stability and evaluation time. 
Our results are implemented in the \texttt{Mathematica} package \texttt{TTH}.}

\allowdisplaybreaks

\begin{document}
%============================================
\preprint{\begin{minipage}[t]{8cm}\begin{flushright} 
OUTP-23-17P \\
TUM-HEP-1487/23
\end{flushright}\end{minipage}}

\maketitle

\flushbottom

%============================================
%	MAIN PART
%============================================

\newpage
\section{Introduction}
After the discovery of the Standard Model (SM) Higgs boson at the Large Hadron Collider (LHC)~\cite{ATLAS:2012yve,CMS:2012qbp}, 
arguably one of the most interesting and pressing search directions is the characterization of its interactions with other SM particles.
In this respect, the Yukawa sector of the SM~\cite{10.1143/PTPS.1.1} plays a special but peculiar role, since the coupling of the Higgs boson
to charged fermions is responsible for their masses. The peculiarity is given by the fact that 
fermions masses have disparate values in the
SM, spanning five orders of magnitude across different generations and within the same isospin doublet. As far as the value of the mass is concerned, 
the clear outlier is the top-quark, with a value of approximately 170 GeV and a Yukawa coupling predicted by the SM to be of $\mathcal{O}(1)$.
Therefore, a deep understanding of the top-Higgs interaction offers a unique path to a potential realization of the SM Yukawa puzzle and can serve as
a portal to physics beyond the SM.

A direct measurement of the top-quark Yukawa coupling is attained in the associated production of a Higgs boson with a top anti-top ($t\bar{t}$) pair.
The discovery of this production channel was first reported by the ATLAS and CMS collaborations~\cite{ATLAS:2018mme,CMS:2018uxb} in 2018, and
since then, great work has been put in better characterizing its properties~\cite{ATLAS:2020ior,CMS:2020cga}.
Both experimental collaborations currently report an accuracy on the signal strength of roughly $\mathcal{O}(15-20\%)$~\cite{Cepeda:2019klc}.
This number is bound to significantly decrease in the High-Luminosity (HL) phase of the LHC. 
This will reduce the relative impact of statistical uncertainty and leave
the theory systematics as the dominant outstanding source of uncertainty, which is currently estimated to be of $\mathcal{O}(10\%)$~\cite{LHCHiggsCrossSectionWorkingGroup:2016ypw,Cepeda:2019klc}. 
Projections for the HL-LHC, quote a total uncertainty of $\mathcal{O}(2\%)$.

These projections call for the most accurate possible theoretical predictions within the SM. 
On the side of fixed-order calculations, the first Leading-Order (LO) studies were published
almost forty years ago~\cite{PhysRevD.29.876,Kunszt:1984ri}, followed, twenty years later, by higher-order QCD corrections at Next-to-Leading (NLO)~\cite{Beenakker:2001rj,Reina:2001sf,Reina:2001bc,Beenakker:2002nc,Dawson:2003zu}. NLO electroweak corrections have also been considered ~\cite{Frixione:2014qaa,Zhang:2014gcy,Frixione:2015zaa}, later including off-shell effects from top-quark decays as well~\cite{Denner:2016wet}.
Important progress has recently been made with the first approximate calculation of the inclusive cross-section 
for $t\bar{t}$ Higgs associated production
through NNLO in QCD~\cite{Catani:2021cbl,Catani:2022mfv}. 
One of the main outcome is a significant reduction of the theoretical uncertainty, ranging in the few percent 
region at current and future colliders. This calculation has been performed retaining all the ingredients necessary for a fully differential description
of this process, except for the genuine two-loop corrections to the matrix element. 
As for the latter, a so-called soft-Higgs approximation is employed, where remarkable simplifications occur due to the factorization of the amplitude in this limit. The accuracy of this approximation is based on the expectation that
two-loop corrections are small in the inclusive rate. 
However, for a formally NNLO accurate prediction, the two-loop corrections is a necessary element and it would be interesting to assess the validity
of the soft-Higgs approximation in extreme kinematic regimes.

The multi-scale nature of this problem renders a full two-loop calculation a formidable task.
Indeed, only recently the computation of all relevant two-loop five-point amplitudes with massless external particles have been completed~\cite{Agarwal:2021vdh,Badger:2021imn,Abreu:2023bdp,Badger:2023mgf,Agarwal:2023suw,DeLaurentis:2023nss,DeLaurentis:2023izi}.
In the case of one external massive particle, first results in leading-color approximation have been computed~\cite{Badger:2021nhg,Abreu:2021asb,Badger:2021ega,Badger:2022ncb}, and although ingredients for going beyond the planar limit are
now available~\cite{Abreu:2023rco} a full amplitude calculation still presents serious challenges.
Due to the rapidly increasing complexity with the number of scales, to this day, no two-loop five-point scattering amplitude with two massive external states has been computed.
Progress in this direction has been made with the calculation of the $t\bar t$ plus one jet one-loop amplitude to higher orders in the dimensional regulator~\cite{Badger:2022mrb} 
and of a set of planar two-loop integrals~\cite{Badger:2022hno}. 
A process like $t\bar{t}$ Higgs, with three massive external states, clearly represents the boundary
of current technology. Nevertheless, the interest in this calculation is also demonstrated by 
first results for some classes of planar two-loop integrals, which
have appeared very recently~\cite{FebresCordero:2023gjh}.

In this paper, we take a preliminary step towards a complete calculation of the two-loop 
$t\bar{t}$ plus Higgs scattering amplitude, by computing 
the one-loop corrections to the  phenomenologically dominant gluon-gluon channel
to higher-orders in the dimensional regulator.
First results for the unpolarized amplitude for this process up to order $\mathcal{O}(\epsilon)$
have recently been obtained in~\cite{Chen:2022nxt}. In that reference, the relevant master integrals
have been identified and a canonical basis~\cite{Henn:2013pwa} has been provided. 
In our paper, we go one step forward and propose an efficient strategy to evaluate both polarized and unpolarized scattering
amplitudes for this processes to order $\mathcal{O}(\epsilon^2)$, which is the order required to
fully define the finite remainder of the corresponding two-loop amplitudes.

The rest of this paper is organized as follows. After reviewing our conventions and color decomposition in Sec.~\ref{sec:conventions}, we discuss the generalization of massless helicity amplitudes within the projector method in Sec.~\ref{sec:Helicity-Chirality_amplitude}. In Sec.~\ref{subsec:renormalization}, we discuss  Ultraviolet (UV) renormalization and  Infrared (IR) subtraction. Using the integral topologies of \cite{Chen:2022nxt}, in Sec.~\ref{sec:details_of_computation} we identify a non-redundant set of master integrals to express the physical amplitudes, exposing extra relations missed by a naive application of standard reduction programs. In Sec.~\ref{sec:analaytic_challenges}, we discuss several approaches to cope with the complexity inherent in a fully analytic calculation, including  the analytic integration of the master integrals and the analytic reduction of the scattering amplitude using Integration-by-Parts (IBP) relations~\cite{Tkachov:1981wb,Chetyrkin:1981qh}. 
We propose an alternative semi-numerical approach in Sec.~\ref{sec:seminumerical} in terms of an augmented version of the Auxiliary Mass Flow algorithm, which allows the evaluation of the scattering amplitude to order $\eps^2$ efficiently and in a stable way.
Our final results are implemented in the proof-of-concept \texttt{Mathematica} package \texttt{TTH}, which can be downloaded from git via
\begin{center}
    \GIT\, .
\end{center}

\section{Conventions, Kinematics and Color}
\label{sec:conventions}
We consider the production of a $t\bar{t}$ pair in association with a Higgs boson $H$ in gluon fusion,
where all particles are on their mass shell, i.e. we do not consider decays of the top quarks and the Higgs boson.
We define the scattering process taking all particles to be incoming,
\begin{equation}
\label{eq:processdef}
      g(p_1) + g(p_2) + t(p_3) + \overline{t}(p_4) + H(p_5) \to 0\,,    
\end{equation}
such that momentum conservation implies
\begin{equation}
    p_1 + p_2 + p_3 + p_4 + p_5 = 0.
    \label{eq:momentum_conservation}
\end{equation}
The on-shell conditions read
\begin{equation}
    p_1^2 = p_2^2 = 0, \;\; p_3^2 = p_4^2 = m_t^2, \;\; \text{ and } \;\; p_5^2 = m_H^2,
    \label{eq:onshell_conditions}
\end{equation}
where $m_t$ denotes the top-quark mass and $m_H$ the Higgs boson mass. 
The other quarks are taken to be massless. 

The kinematics of the process is described through the Mandelstam variables $s_{ij}=(p_i+p_j)^2$. Using momentum conservation~\eqref{eq:momentum_conservation} and 
the on-shell conditions~\eqref{eq:onshell_conditions}, we relate all kinematic invariants to the minimal set
\begin{equation}\label{eq:mandelstam}
    \{s_{12}, s_{13}, s_{14}, s_{23}, s_{24}, s_{34}, m_t^2\}\,.
\end{equation}
This set of variables is closed under the exchanges of the momenta $p_1\leftrightarrow p_2$ and $p_3\leftrightarrow p_4$. We will exploit this feature later in our discussion. 
Unless stated otherwise, we rescale all Mandelstam variables by $m_t^2$, which is equivalent to setting $m_t=1$.
Furthermore, we define the Gram determinant built out of the four independent momenta $\left\lbrace p_1,\ldots,p_4\right\rbrace$ as
\begin{equation}\label{eq:defgram}
    \Delta \equiv 
    \operatorname{det} \left(  p_i \cdot p_j
    \right) = G(p_1,p_2,p_3,p_4).  
\end{equation}
In the physical scattering region, $g\,g \,\to \,\bar{t}\,t\, H$, one has $\Delta < 0$~\cite{RevModPhys.36.595}. 
For the description of polarized amplitudes, one further needs the parity-odd quantity
\begin{equation}
    {\rm tr}_5 \equiv i \varepsilon^{p_1p_2p_3p_4}\equiv i\varepsilon_{\mu_1\mu_2\mu_3\mu_4} 
    p_1^{\mu_1} p_2^{\mu_2} p_3^{\mu_3} p_4^{\mu_4}\,,
    \label{eq:relation_tr5_epsilon}
\end{equation}
which is related to the Gram determinant through
\begin{equation}
   \Delta =  {\rm tr}_5^2.
    \label{eq:def_tr5_GRAM}
\end{equation}

Let us now discuss the general structure of the scattering amplitude for the process in Eq.~\eqref{eq:processdef}.
It can be expressed as
\begin{equation} \label{eq:expA}
    \mathcal{A} = y^0_t\left(4 \pi \alpha^0_s\right) \, \sum_{\ell=0}^\infty
    \left(\frac{\alpha_s^0}{4\pi}\right)^{\ell} \mathcal{A}^{(\ell)}\,,
\end{equation}
where we perturbatively expand in powers of the bare strong-coupling constant $\alpha_s^0$, 
and where $\ell$ refers to the $\ell$-loop contribution. We also factor out
the leading-order terms, $\ell=0$, in $\alpha_s^0$ and in the bare top-quark Yukawa coupling $y^0_t$.
The latter is defined via the bare top-quark mass $m^0_t$ and the vacuum-expectation value $v$ as
\begin{equation}
  y^0_t = \frac{m^0_t}{v}\,.
\end{equation}

At any loop order, the scattering amplitude can be further decomposed into three gauge-independent color structures,
\begin{equation}\label{eq:colorF}
    \mathcal{A}^{(\ell)} = \mathcal{A}^{(\ell)}_1 \ket{\mathcal{C}_1} 
    +  \mathcal{A}^{(\ell)}_2 \ket{\mathcal{C}_2}
    +  \mathcal{A}^{(\ell)}_3 \ket{\mathcal{C}_3}\,,
\end{equation}
where $\ket{\mathcal{C}_{i}}$ are basis elements of the color vector space and $\mathcal{A}^{(\ell)}_{i}$ are the so-called
partial amplitudes. For the color basis we choose
\begin{equation}
    \label{eq:def_color_basis}
    \ket{\mathcal{C}_1} \equiv  T_{i_4 k}^{a_1}T_{k i_3}^{a_2}, 
    \quad \quad \ket{\mathcal{C}_2} \equiv T_{i_4 k}^{a_2}T_{k i_3}^{a_1}, 
    \quad \quad \ket{\mathcal{C}_3} \equiv \delta^{a_1 a_2}\delta_{i_4 i_3}\,.
\end{equation}
Here, $a_n$ are indices in the adjoint representation and they refer to the gluons, 
whereas $i_n$ are in the fundamental representation and they refer to the top-quarks. 
The color operators $T^{a}_{ij}$ satisfy the normalization condition $\mathrm{Tr}\left[T^{a}T^{b}\right] = \delta^{ab}/2$. 
Let us note that the coefficient $\mathcal{A}^{(0)}_3$ is identically zero, 
thus at leading-order only $\ket{\mathcal{C}_{1,2}}$ contribute.
\section{Helicity-Chirality Amplitudes in the Projector Method}
\label{sec:Helicity-Chirality_amplitude}

Besides the color-decomposition, the scattering amplitude also admits a decomposition into Lorentz structures, often referred to as tensors
 $T_i$. 
The latter multiply coefficients, so-called form factors $F_i$, which transform trivially under the action of the Lorentz group
\begin{equation}
    \mathcal{A}= \sum_i F_i\, T_i\,. 
    \label{eq:AmplitudeInTensors}
\end{equation}
While the tensors are loop independent, the form factors $F_i$ are not. 
As in Eq.~\eqref{eq:expA}, we expand the $F_i$ in powers of $\alpha^0_s$, 
so that they are related to the perturbative coefficients $\mathcal{A}^{(\ell)}$ via
\begin{equation}\label{eq:expF}
    \mathcal{A}^{(\ell)} = F^{(\ell)}_i T_i\, .
\end{equation}
This decomposition is valid for any partial amplitude $\mathcal{A}^{(\ell)}_j$ and is carried out independently of color. 
Therefore, to ease readability, we suppress color indices in the following discussion.

In the 't Hooft-Veltman (tHV) dimensional regularization scheme~\cite{tHooft:1972tcz}, 
where external states are four dimensional, 
the number of independent tensor structures is in one-to-one correspondence with the number of 
helicity configurations of the external particles~\cite{Peraro:2019cjj,Peraro:2020sfm}. 
In our case, we have two massless spin-$1$ bosons and two massive spin-$1/2$ fermions, which account in total for $2 \time 2 \times 2 \times 2 \times 2 = 16$ different polarizations.

A spanning basis for the corresponding tensor structures is given by
\begin{equation}
    t _{ijk} = \bar{v}(p_4) \Gamma_i u(p_3) \, 
    \varepsilon_1 \cdot p_j \, \varepsilon_2 \cdot p_k\,, \quad \mbox{with} \quad j,k = 3,4\,.\label{eq:tensors}
\end{equation}
Here, $\bar{v}(p_4)$ and $u(p_3)$ are the four-dimensional spinors of the antitop and the top quarks, $\varepsilon_1^\mu$ and $\varepsilon_2^\mu$ are the 
four-dimensional polarization vectors of the two gluons, and 
\begin{equation}
    \Gamma_i = \{\mathbb{I}, \slashed{p}_1, \slashed{p}_2, \slashed{p}_1\slashed{p}_2  \}\,. \label{eq:Gammaset}
\end{equation}

We derived the tensor basis by noticing that the four independent momenta $p_1^\mu,p_2^\mu,p_3^\mu,p_4^\mu$ span the whole four-dimensional space. 
Hence, any four-dimensional Lorentz tensor can be expressed in terms of these momenta. This is true in particular for the $\gamma$-matrices
\begin{equation}
    \gamma^\mu = \sum_{i=1}^4 \hat{a}_i p_i^{\mu}\,,
\end{equation}
where the coefficients $\hat{a}_i$ are operators in spinor space built out of $\slashed{p}_i$. 
Their explicit form is immaterial for the derivation of a spanning tensor basis, as 
they can be absorbed in the overall normalization of the tensors. Hence, $\Gamma_i$ is obtained 
by inserting the $\slashed{p}_i$ in all possible ways and noticing that
combinations with three or more instances of $\slashed{p}_i$ are linked to the previous ones through Dirac algebra.
Importantly, due to the Dirac equation
\begin{equation}
(\slashed{p}_3-m) u(p_3) = 0\,, \quad \bar{v}(p_4) (\slashed{p}_4 +m )= 0\,,
\label{eq:Dirac_Eq}
\end{equation}
$\Gamma_i$ cannot have any dependence on $\slashed{p}_3$ and $\slashed{p}_4$.
Finally, we impose transversality on the external gluons
\begin{equation}
    \label{eq:physstates}
    \varepsilon_1 \cdot p_1 = \varepsilon_2 \cdot p_2 = 0,    
\end{equation}
and choose the gluons' reference vectors such that
\begin{equation}
    \varepsilon_1 \cdot p_2 = \varepsilon_2 \cdot p_1 = 0\,.
    \label{eq:gaugech}
\end{equation}
%
%Note that our choice of reference vectors is symmetric under the exchange of the two gluons. 
These constraints leave us with the spanning basis in Eq.~\eqref{eq:tensors}.

As expected, there are $16$ independent tensors. There is clearly some freedom in the choice of a basis, and the latter has a significant impact on the complexity of the corresponding form factors. For convenience, we choose tensors which are either 
symmetric or anti-symmetric under the exchange of the two gluons. We further order the tensors into two groups, the first one consisting of $4$ and the second one of $12$ tensors. Explicitly:

\allowdisplaybreaks
\subsubsection*{Group 1:}
\begin{align}
\mathbf{1.}\quad T_{1S} &= m_t \Big[\big(\varepsilon_1\cdot p_3\varepsilon_2\cdot p_4 - \varepsilon_1\cdot p_4\varepsilon_2\cdot p_3\big)\overline{v}(p_4)\big(\!\not\!{p}_1 - \!\not\!{p}_2\big)u(p_3)\Big] \notag,\\
\mathbf{2.}\quad T_{2S} &=  \Big[\big(\varepsilon_1\cdot p_3\varepsilon_2\cdot p_4 - \varepsilon_1\cdot p_4\varepsilon_2\cdot p_3\big)\overline{v}(p_4)\big(\!\not\!{p}_1\!\not\!{p}_2 - \!\not\!{p}_2\!\not\!{p}_1\big)u(p_3)\Big] \notag,\\
\mathbf{3.}\quad T_{3A}  &= m_t^2\Big[\big(\varepsilon_1\cdot p_3\varepsilon_2\cdot p_4 - \varepsilon_1\cdot p_4\varepsilon_2\cdot p_3\big)\overline{v}(p_4)u(p_3)\Big] \notag,\\
\mathbf{4.}\quad T_{1A}  &= m_t \Big[\big(\varepsilon_1\cdot p_3\varepsilon_2\cdot p_4 - \varepsilon_1\cdot p_4\varepsilon_2\cdot p_3\big)\overline{v}(p_4)\big(\!\not\!{p}_1 + \!\not\!{p}_2\big)u(p_3)\Big],
\label{eq:def_tensors_group1}
\end{align}
\subsubsection*{Group 2:}
\begin{align}
    \mathbf{5.}\quad T_{3S}  &= m_t^2\Big[\big(\varepsilon_1\cdot p_3\varepsilon_2\cdot p_4 + \varepsilon_1\cdot p_4\varepsilon_2\cdot p_3\big)\overline{v}(p_4)u(p_3)\Big] \notag,\\
    \mathbf{6.}\quad T_{4S}  &= m_t \Big[\varepsilon_1\cdot p_3\varepsilon_2\cdot p_3\overline{v}(p_4)\big(\!\not\!{p}_1 + \!\not\!{p}_2\big)u(p_3)\Big]  \notag,\\
    \mathbf{7.}\quad T_{5S}  &= m_t \Big[\varepsilon_1\cdot p_4\varepsilon_2\cdot p_4\overline{v}(p_4)\big(\!\not\!{p}_1 + \!\not\!{p}_2\big)u(p_3)\Big]  \notag,\\
    \mathbf{8.}\quad T_{6S}  &= m_t \Big[\big(\varepsilon_1\cdot p_3\varepsilon_2\cdot p_4 + \varepsilon_1\cdot p_4\varepsilon_2\cdot p_3\big)\overline{v}(p_4)\big(\!\not\!{p}_1 + \!\not\!{p}_2\big)u(p_3)\Big], \notag\\
    \mathbf{9.}\quad T_{7S}  &=  \Big[\varepsilon_1\cdot p_3\varepsilon_2\cdot p_3\overline{v}(p_4)\big(\!\not\!{p}_1\!\not\!{p}_2 + \!\not\!{p}_2\!\not\!{p}_1\big)u(p_3)\Big] \notag\\
     &= s_{12}\Big[\varepsilon_1\cdot p_3\varepsilon_2\cdot p_3\overline{v}(p_4)u(p_3)\Big],  \notag\\
    \mathbf{10.}\quad T_{8S} &= \Big[\varepsilon_1\cdot p_4\varepsilon_2\cdot p_4\overline{v}(p_4)\big(\!\not\!{p}_1\!\not\!{p}_2 + \!\not\!{p}_2\!\not\!{p}_1\big)u(p_3)\Big] \notag\\
    &= s_{12}\Big[\varepsilon_1\cdot p_4\varepsilon_2\cdot p_4\overline{v}(p_4)u(p_3)\Big]  \notag,\\
   \mathbf{11.}\quad  T_{4A}  &= m_t \Big[\varepsilon_1\cdot p_3\varepsilon_2\cdot p_3\overline{v}(p_4)\big(\!\not\!{p}_1 - \!\not\!{p}_2\big)u(p_3)\Big]  \notag,\\
\mathbf{12.}\quad T_{5A} &= m_t \Big[\varepsilon_1\cdot p_4\varepsilon_2\cdot p_4\overline{v}(p_4)\big(\!\not\!{p}_1 - \!\not\!{p}_2\big)u(p_3)\Big]  \notag,\\
\mathbf{13.}\quad T_{6A} &= m_t \Big[\big(\varepsilon_1\cdot p_3\varepsilon_2\cdot p_4 + \varepsilon_1\cdot p_4\varepsilon_2\cdot p_3\big)\overline{v}(p_4)\big(\!\not\!{p}_1 - \!\not\!{p}_2\big)u(p_3)\Big] \notag,\\
\mathbf{14.}\quad T_{7A} &=  \Big[\varepsilon_1\cdot p_3\varepsilon_2\cdot p_3\overline{v}(p_4)\big(\!\not\!{p}_1\!\not\!{p}_2 - \!\not\!{p}_2\!\not\!{p}_1\big)u(p_3)\Big]  \notag,\\
\mathbf{15.}\quad T_{8A} &=  \Big[\varepsilon_1\cdot p_4\varepsilon_2\cdot p_4\overline{v}(p_4)\big(\!\not\!{p}_1\!\not\!{p}_2 - \!\not\!{p}_2\!\not\!{p}_1\big)u(p_3)\Big] \notag ,\\
\mathbf{16.}\quad T_{2A} &=  \Big[\big(\varepsilon_1\cdot p_3\varepsilon_2\cdot p_4 + \varepsilon_1\cdot p_4\varepsilon_2\cdot p_3\big)\overline{v}(p_4)\big(\!\not\!{p}_1\!\not\!{p}_2 - \!\not\!{p}_2\!\not\!{p}_1\big)u(p_3)\Big].
\label{eq:def_tensors_group2}
\end{align}
Tensors belonging to distinct groups are mutually orthogonal, i.e. 
\begin{equation}
     T_i^{\dagger} \cdot T_j = 0, 
    \quad \mbox{if} \;\; i=1,...,4;\;\; j=5,...,16\,,
\end{equation}
where $T^{\dagger}_i$ are the dual tensors. The scalar product among 
tensors and their dual ones is defined by summing over polarization of the 
external particles
\begin{equation}
    T_i^\dagger \cdot T_j = \sum_{\mathrm{pol}} T_i^\dagger T_j \,. \label{eq:scalprod}
\end{equation}
For consistency with our choice of reference vectors Eq.~\eqref{eq:gaugech}, one must use the polarization sum rule
\begin{equation}
    \sum_{pol} \varepsilon_1^\mu \varepsilon_1^{\nu*} = \sum_{pol} \varepsilon_2^\mu \varepsilon_2^{\nu*} = -g^{\mu \nu}
    + \frac{p_1^\mu p_2^\nu + p_2^\mu p_1^\nu}{p_1\cdot p_2}\,. 
    \label{eq:polsum}
\end{equation}

Next, we construct projector operators to single out the individual form factors
\begin{equation}
    P_i  = \sum_{j=1}^{16} \left(M^{-1}\right)_{ij}T_j, \quad\text{ with }\quad M_{ij} = T^{\dagger}_i\cdot T_j.
    \label{eq:def_projector}
\end{equation}
The matrix $M^{-1}$ contains in general inverse powers of the Gram determinant $\Delta$, Eq.~\eqref{eq:defgram}. This is expected, since it follows from
the fact that the tensors become linearly dependent if four external momenta are not all independent.
For our specific tensor choice~\eqref{eq:tensors}, the inverse matrix $M^{-1}$ contains a global factor of  $\Delta^{-2}$ for the tensors belonging to the first group, and $\Delta^{-3}$ for the tensors belonging to the second group. Interestingly, this dependence is milder than a generic tensor basis choice in which an overall factor $\Delta^{-3}$ multiplies all tensors.

At tree level, these inverse powers of $\Delta$ are clearly a spurious residue of the projector method, 
since none of the individual Feynman diagrams depend on $\Delta$. After an explicit computation, we verified that the tree level form factors associated to these tensors, contain indeed inverse powers of $\Delta$. 
As a matter of fact, in massless multileg calculations,
the residual $\Delta$ dependence cancels out after recombining the form factors to form physical helicity amplitudes. 
We expect similar simplifications in the presence of massive external states, 
if one recombines the form factor into physical quantities.

The generalization, however, is non-trivial. 
In fact, for massless particles, helicity is a well defined quantum number and it is convenient to represent helicity amplitudes using massless spinor helicity formalism, see e.g.~\cite{Dixon:1996wi}. In contrast, for massive particles, helicity is a frame dependent quantum number. While spinor helicity formalism can be generalized to the massive case, 
see for example~\cite{Arkani-Hamed:2017jhn,Badger:2021owl} and references therein, it is not obvious that decomposing the amplitude in helicity eigenstates is the right thing to do.
In our calculation, we decide to follow an hybrid approach, which allows us to see the explicit cancellation of the
the unphysical Gram determinant $\Delta$ at tree-level, without committing to a specific choice of massive spinor helicity formalism.

Gluons are massless and therefore helicity is a good quantum number to label their quantum states. 
There are four configurations for the two gluons
$(+,+)$, $(+,-)$, $(-,+)$, $(-,-)$ of which two are related by parity. We choose $(+,+)$ and $(+,-)$ as independent helicities and  fix them explicitly by rewriting the polarization vectors $\varepsilon_1$ and $\varepsilon_2$ in spinor helicity formalism 
\begin{align}
    \varepsilon^{\mu}_{1+} = \frac{1}{\sqrt{2}}\frac{\langle 2 \gamma^{\mu} 1 ]}{\braket{12}}, \quad \varepsilon^{\mu}_{1-} = \frac{1}{\sqrt{2}}\frac{\langle1 \gamma^{\mu} 2 ]}{[12]}, \notag\\
    \varepsilon^{\mu}_{2+} = \frac{1}{\sqrt{2}}\frac{\langle 1 \gamma^{\mu} 2 ]}{\braket{21}}, \quad \varepsilon^{\mu}_{2-} = \frac{1}{\sqrt{2}}\frac{\langle 2 \gamma^{\mu} 1 ]}{[21]}.
\end{align}
The choice of reference vectors corresponds to the choice in Eq.~\eqref{eq:gaugech}.

Let us consider the two independent possibilities separately.
In the $(+,+)$ configuration, it is easy to rewrite the product of the two polarizations vectors for the two gluons as a trace 
\begin{align}
\varepsilon_{1+}^{\mu}\varepsilon_{2+}^{\nu} = -\dfrac{1}{2\left\langle12\right\rangle^2} \left\langle 2\left|\gamma^{\mu}\right| 1\right]\left\langle 1\left|\gamma^{\nu}\right| 2\right] 
= \Phi_{++}\operatorname{tr}\left[P_L\not\!p_2\gamma^{\mu}\!\!\not\!p_1\gamma^{\nu}\right],
\label{eq:pp_dirac_tr}
\end{align}
where the overall spinor weight is collected out in the spinor phase $\Phi_{++}$
\begin{equation}
    \Phi_{++} \equiv -\dfrac{1}{2\left\langle12\right\rangle^2}.
    \label{eq:spinor_phase_pp}
\end{equation}
The polarization vectors appear in the various tensor structures contracted with $p_3^\mu$ and $p_4^\mu$. 
More precisely, we require the following four contractions
\begin{align}
    c_{3,3}^{++} &\equiv \Phi^{-1}_{++} \left[p_3\cdot \varepsilon^{(+)}_{1}p_3\cdot \varepsilon^{(+)}_{2}\right]  = m_t^4-m_t^2 (s_{12}+s_{13}+s_{23})+s_{13} s_{23}, \notag\\
    c_{4,4}^{++} &\equiv \Phi^{-1}_{++} \left[ p_4\cdot \varepsilon^{(+)}_{1}p_4\cdot \varepsilon^{(+)}_{2} \right] = m_t^4-m_t^2 (s_{12}+s_{14}+s_{24})+s_{14} s_{24}, \notag\\
    c_{3,4;\,S}^{++} &\equiv \Phi^{-1}_{++} \left[ p_3\cdot \varepsilon^{(+)}_{1}p_4\cdot \varepsilon^{(+)}_{2} + p_4\cdot \varepsilon^{(+)}_{1}p_3\cdot \varepsilon^{(+)}_{2}\right]
    \notag\\&= m_t^2 (2 s_{12}-s_{13}-s_{14}-s_{23}-s_{24}) -s_{12} s_{34}+s_{13} s_{24}+s_{14} s_{23} + 2 m_t^4, \notag\\
    c_{3,4;\,A}^{++} &\equiv \Phi^{-1}_{++} \left[  p_3\cdot \varepsilon^{(+)}_{1}p_4\cdot \varepsilon^{(+)}_{2} - p_4\cdot \varepsilon^{(+)}_{1}p_3\cdot \varepsilon^{(+)}_{2}\right] = -4 \,{\rm tr}_5.
    \label{eq:pp_contractions}
\end{align}

We repeat the same steps for the $(+,-)$ configuration. The only difference is that we must introduce 
two auxiliary vectors $r, q$ in order to close the Dirac trace and to extract a spinor phase. In particular, we write
\begin{align}
\varepsilon_{1+}^{\mu}\varepsilon_{2-}^{\nu}   &= 
-\dfrac{1}{2s_{12}} \langle 2 \gamma^{\mu} 1 ]
\langle 2 \gamma^{\nu} 1 ] \notag\\
&= -\dfrac{1}{2s_{12}} \frac{ \langle 2 \gamma^{\mu} 1 ] \langle 1 \,r\,  2 ]
\langle 2 \gamma^{\nu} 1 ] \langle 1 \,q\, 2]}{\langle 1\,r\, 2 ] \langle 1\,q\, 2 ]} \notag\\
&= -\dfrac{1}{2s_{12}}\frac{\operatorname{tr} \left[P_L\not\!p_2\gamma^{\mu}\!\! \not \!p_1\!\! \not \!r\!\! \not \!p_2 \gamma^{\nu} \!\! \not\!p_1 \!\!\not \!q\right]}{ \langle 1\,r\, 2 ] 
\langle 1\, q\, 2]}
 \notag\\
&= \Phi_{+-}\operatorname{tr}\left[P_L\not\!p_2\gamma^{\mu} \!\!\not\!p_1\!\!\not\!r\!\!\not\!p_2\gamma^{\nu}\!\!\not\!p_1\!\!\not\!q\right]\,,
\label{eq:pm_dirac_tr}
\end{align}
with
\begin{equation}
    \Phi_{+-} \equiv -\dfrac{1}{2s_{12}}\frac{1}{\langle 1\,r\, 2 ] \langle 1\,q\, 2]}.  
\label{eq:spinor_phase_pm}
\end{equation}
In order to preserve the symmetry under the exchange $\mu\leftrightarrow\nu$, it is convenient to choose $r=q=p_3$. 
Contracting Eq.~\eqref{eq:spinor_phase_pm} with $p_3$ and $p_4$ yields
\begin{align}
    c_{3,3}^{+-} &\equiv \Phi^{-1}_{+-}\left[ p_3\cdot \varepsilon^{(+)}_{1}p_3\cdot \varepsilon^{(-)}_{2} \right]  = \frac{1}{m_t^2}\left(m_t^4-m_t^2 (s_{12}+s_{13}+s_{23})+s_{13} s_{23}\right)^2, \notag\\
    c_{4,4}^{+-} &\equiv \Phi^{-1}_{+-} \left[p_4\cdot \varepsilon^{(+)}_{1}p_4\cdot \varepsilon^{(-)}_{2} \right] = \hat{c}_{4,4}^{+-} + tr_5 \Tilde{c}_{4,4}^{+-}, \notag\\
    c_{3,4;\,S}^{+-} &\equiv \Phi^{-1}_{+-} \left[p_3\cdot \varepsilon^{(+)}_{1}p_4\cdot \varepsilon^{(-)}_{2} + p_4\cdot \varepsilon^{(+)}_{1}p_3\cdot \varepsilon^{(-)}_{2}\right] = \hat{c}_{3,4;\,S}^{+-} + tr_5 \Tilde{c}_{3,4;\,S}^{+-},\notag\\
    c_{3,4;\,A}^{+-} &\equiv  
     \Phi^{-1}_{+-} \left[p_3\cdot \varepsilon^{(+)}_{1}p_4\cdot \varepsilon^{(-)}_{2} - p_4\cdot \varepsilon^{(+)}_{1}p_3\cdot \varepsilon^{(-)}_{2}\right]  = 0,
    \label{eq:pm_contactions}
\end{align}
with 
\begin{align}
\Tilde{c}_{4,4}^{+-} &= \frac{2}{m_t^2} \left[ m_t^4+m_t^2 ( s_{12}-s_{13}-s_{14})-\frac{1}{2}s_{12} s_{34}+s_{13} s_{24} + \{1\leftrightarrow 2\}\right],\notag\\
    \hat{c}_{4,4}^{+-} &= \frac{1}{2}\bigg\{m_t^2 \Big[s_{12}^2-6 s_{12} (s_{13}+s_{14})-2 s_{12} s_{34}+s_{13}^2+4 s_{13} s_{24}+s_{14}^2\Big]\notag\\
    &+2m_t^4(3 s_{12}-s_{13}-s_{14}) + \left[-2 s_{12}^2 s_{34}-2 s_{13} s_{24} (s_{13}+s_{24})\right.\notag\\
    &+s_{12} (2 s_{13}s_{34}+2 s_{14} s_{34}+s_{13} s_{23}+4 s_{13} s_{24}+s_{14} s_{24})\Big] + \{1\leftrightarrow2\} \bigg\}  \notag\\
    &+ \frac{1}{2m_t^2}\left[(s_{13} s_{24}-s_{12} s_{34})^2-2 s_{12} s_{14} s_{23} s_{34}+s_{14}^2 s_{23}^2 + 2 m_t^8\right],\notag\\
    \Tilde{c}_{3,4;\,S}^{+-} &= 4\left(m_t^4-m_t^2 (s_{12}+s_{13}+s_{23})+s_{13} s_{23}\right), \notag\\
    \hat{c}_{3,4;\,S}^{+-}   &= \frac{1}{4}\Tilde{c}_{3,4;\,S}^{+-}\left[2 m_t^4 - s_{12} s_{34} + [m_t^2 (s_{12}-s_{13}-s_{14})+s_{13} s_{24}] + (1\leftrightarrow2)\right].
\end{align}

With these results, we now rotate our original tensor basis to the new one $\mathcal{T}_j$
\begin{align}
    \begin{array}{llll}
        \mathcal{T}_1 = \Phi_{++}\Psi_1, &\;\; 
        \mathcal{T}_2 = \Phi_{++}\Psi_2,  & \;\;
        \mathcal{T}_3 = \Phi_{++}\Psi_3, &\;\; 
        \mathcal{T}_4 = \Phi_{++}\Psi_4, \\
        \mathcal{T}_5 = \Phi_{+-}\Psi_1, &\;\; 
        \mathcal{T}_6 = \Phi_{+-}\Psi_2,  &\;\; 
        \mathcal{T}_7 = \Phi_{+-}\Psi_3, &\;\; 
        \mathcal{T}_8 = \Phi_{+-}\Psi_4\,,
    \end{array}
    \label{eq:def_helicity_tensor_basis}
\end{align}
where we introduced the combinations of spinor spinor structures
\begin{align}
    \Psi_{1} &\equiv s_{12}\overline{v}(p_4)u(p_3), \qquad
    \Psi_{2} \equiv \overline{v}(p_4)\big(\!\not\!{p}_1\!\not\!{p}_2 - \!\not\!{p}_2\!\not\!{p}_1\big)u(p_3), \notag\\
    \Psi_{3} &\equiv m_t\overline{v}(p_4)\big(\!\not\!{p}_1 + \!\not\!{p}_2\big)u(p_3), \qquad
    \Psi_{4} \equiv m_t\overline{v}(p_4)\big(\!\not\!{p}_1 - \!\not\!{p}_2\big)u(p_3). 
\label{eq:psi_chirality}
\end{align}
The four structures in Eq.~\eqref{eq:psi_chirality} are independent in $D=4$ space-time dimensions, as one can check by verifying that
their Gram matrix
has full rank
\begin{equation}
    \operatorname{det}(\chi_{ij}) \neq 0\,,\quad \text{ with } \quad \chi_{ij}\equiv\sum\limits_{\rm spin}\Psi_i^{\dagger}\Psi_j.
    \label{eq:def_chimatrix}
\end{equation}

We refer to the form factors  $\mathcal{F}_i$ corresponding to the new tensor basis $\mathcal{T}$ as \emph{helicity form factors}. We decompose them into an even and an odd part under the action of parity transformations
\begin{equation}
    \mathcal{F}_i = \mathcal{F}_i^{\text{even}} + \tr_5\,\mathcal{F}_i^{\text{odd}}.
    \label{eq:helicity_formfactors_even_odd}
\end{equation}
With these, the four helicity amplitudes with fixed gluon helicities become explicitly
\begin{align}
    \mathcal{A}^{(\ell)}_{++} &\equiv \Phi_{++}\sum\limits_{i=1}^4 \left(\mathcal{F}^{(\ell)\text{even}}_{i} + \tr_5 \mathcal{F}^{(\ell)\text{odd}}_{i}\right) \Psi_{i}, \notag\\
    \mathcal{A}^{(\ell)}_{+-} &\equiv \Phi_{+-}\sum\limits_{i=5}^8 \left(\mathcal{F}^{(\ell)\text{even}}_{i} + \tr_5 \mathcal{F}^{(\ell)\text{odd}}_{i}\right) \Psi_{(i-4)}, \notag\\
    \mathcal{A}^{(\ell)}_{--} &\equiv \Phi_{++}^{\dagger}\sum\limits_{i=1}^4 \left(\mathcal{F}^{(\ell)\text{even}}_{i} - \tr_5 \mathcal{F}^{(\ell)\text{odd}}_{i}\right) \Psi_{i}, \notag\\
    \mathcal{A}^{(\ell)}_{-+} &\equiv \Phi_{+-}^{\dagger}\sum\limits_{i=5}^8 \left(\mathcal{F}^{(\ell)\text{even}}_{i} - \tr_5 \mathcal{F}^{(\ell)\text{odd}}_{i}\right) \Psi_{(i-4)}\,,
    \label{eq:def_helicity_amplitude}
\end{align}
where the relation between the form factors $F_i$ and the helicity form factors $\mathcal{F}_i$ is 
\begin{align}
    \mathcal{F}^{\text{even}}_i = \left(
    \begin{array}{c}
         c^{++}_{3,3}F_{7S} + c^{++}_{4,4}F_{8S} + c^{++}_{3,4;\,S}F_{3S} \\
         c^{++}_{3,3}F_{7A} + c^{++}_{4,4}F_{8A} + c^{++}_{3,4;\,S}F_{2A} \\
         c^{++}_{3,3}F_{4S} + c^{++}_{4,4}F_{5S} + c^{++}_{3,4;\,S}F_{6S} \\
         c^{++}_{3,3}F_{4A} + c^{++}_{4,4}F_{5A} + c^{++}_{3,4;\,S}F_{6A} \\
         c^{+-}_{3,3}F_{7S} + \hat{c}^{+-}_{4,4}F_{8S} + \hat{c}^{+-}_{3,4;\,S}F_{3S} \\
         c^{+-}_{3,3}F_{7A} + \hat{c}^{+-}_{4,4}F_{8A} + \hat{c}^{+-}_{3,4;\,S}F_{2A} \\
         c^{+-}_{3,3}F_{4S} + \hat{c}^{+-}_{4,4}F_{5S} + \hat{c}^{+-}_{3,4;\,S}F_{6S} \\
         c^{+-}_{3,3}F_{4A} + \hat{c}^{+-}_{4,4}F_{5A} + \hat{c}^{+-}_{3,4;\,S}F_{6A} 
    \end{array}
    \right), \quad 
    \mathcal{F}^{\text{odd}}_i = \left(
    \begin{array}{c} 
         \tr_5^{-1}\,c^{++}_{3,4;\,A}F_{3A} \\
         \tr_5^{-1}\,c^{++}_{3,4;\,A}F_{2S} \\
         \tr_5^{-1}\,c^{++}_{3,4;\,A}F_{1A} \\
         \tr_5^{-1}\,c^{++}_{3,4;\,A}F_{1S} \\
         \Tilde{c}^{+-}_{4,4}F_{8S} + \Tilde{c}^{+-}_{3,4;\,S}F_{3S} \\
         \Tilde{c}^{+-}_{4,4}F_{8A} + \Tilde{c}^{+-}_{3,4;\,S}F_{2A} \\
         \Tilde{c}^{+-}_{4,4}F_{5S} + \Tilde{c}^{+-}_{3,4;\,S}F_{6S} \\
         \Tilde{c}^{+-}_{4,4}F_{5A} + \Tilde{c}^{+-}_{3,4;\,S}F_{6A} \\
    \end{array}
    \right).
    \label{eq:linear_comb_hel_form_factors}
\end{align}

A similar decomposition would be desirable for the massive quarks. 
A common way to generalize  spinor helicity formalism for massive particles is to
split each massive momenta into two massless momenta, which are then treated with the conventional spinor helicity formalism. The splitting is arbitrary and necessarily introduces an ambiguity in the calculation. 
Alternatively, the authors of~\cite{Arkani-Hamed:2017jhn} suggested a generalization which manifests the little group scaling of the scattering amplitude. In our case, following the second approach is equivalent to a simple renaming
of the same tensor structures, without any obvious simplifications. While this indicates that the tensor basis in Eq.~\eqref{eq:def_helicity_tensor_basis} is already in a minimal form, we stress that the spinor helicity formalism obscures momentum conservation, such that additional non-trivial rearrangements of the tensor structures cannot be excluded. 
Therefore, we decide to work with our helicity form factors without any further rearrangement. As we will see below, this is sufficient
to guarantee that unphysical powers of the inverse Gram determinant cancel out in the tree level, as expected, and also in many of the one-loop master integral coefficients.
\section{Renormalization and Infrared Structure}
\label{subsec:renormalization}
Our goal is to evaluate the helicity form factors in Eq.~\eqref{eq:def_helicity_amplitude} up to one-loop order.
The results will contain divergences 
both of ultraviolet (UV) and infrared (IR) origin, which we regulate in dimensional regularization.
UV divergences are removed by standard renormalization.
In particular, in our calculation we renormalize the strong-coupling constant in the $\overline{\mathrm{MS}}$ scheme,
whereas we renormalize the top-quark mass, the Yukawa coupling and both the top-quark and gluon wave functions 
in the on-shell scheme, see e.g.~\cite{Denner:2019vbn}.
For later convenience, we introduce the normalization factor
\begin{equation}
    C_{\epsilon} = (4 \pi)^{\epsilon}\,\Gamma(1+\epsilon)\,.
\end{equation}
The renormalized strong-coupling constant $\alpha_s$ is related to the bare one through
\begin{align}
    C_{\epsilon} \, \mu_0^{2\epsilon} \alpha^0_s  = \mu^{2\epsilon} \alpha_s(\mu^2) Z_{\alpha_s} = 
    \mu^{2\epsilon} \alpha_s(\mu^2)\left(1 - \frac{\alpha_s(\mu^2)}{4 \pi} \frac{\beta_0}{\epsilon} +\mathcal{O}(\alpha_s^2)\right),
\end{align}
where $\beta_0$ is the first coefficient of the QCD $\beta$-function,
\begin{equation}
   \beta_0 = \frac{11}{3}C_A-\frac{2}{3}(N_f+N_h)\,,
\end{equation}
with $N_f$ the number of light quarks and $N_h$ the number of heavy quarks, so in our case $N_f = 5$ and $N_h = 1$. 
For definiteness, from now on we fix the renormalization scale to be $\mu=m_t$ and we drop the explicit $\mu$ dependence in
$\alpha_s$. Results for a generic value of $\mu$ can be easily recovered via renormalization group evolution arguments.

The relation between the bare mass $m^0_t$ and the on-shell renormalized mass $m_t$ is~\cite{Melnikov:2000zc}
\begin{equation}
    m^0_t = Z_{m_t} m_t = m_t \left(1 - \frac{\alpha_s}{4\pi}\frac{\delta_{m_t}}{\epsilon} + \mathcal{O}(\alpha_s^2) \right),
\end{equation}
where we introduced
\begin{equation}
\label{eq:topmassct}
    \delta_{m_t} = C_F \left(3 + \frac{4\epsilon}{1-2 \epsilon}\right).
\end{equation}
The one-loop mass renormalization is equivalent to a counter-term insertion into the top-quark propagators,
which is effectively given by
\begin{align}
\frac{i}{\not{\!p}-m^0_t} \, \to \, \frac{i}{\not{\!p}-m_t}\left(1- \frac{\alpha_s}{4\pi}\frac{\delta_{m_t}}{\epsilon}\,\frac{m_t}{\not{\!p}-m_t}\right) + \mathcal{O}\left(\alpha_s^2\right)\,.
\label{eq:Mass_Counterterm}
\end{align}
The renormalization of the top Yukawa coupling is linked to the mass-renormalization simply via
\begin{equation}
    y^0_t = \frac{m^0_t}{v} = \frac{Z_{m_t} m_t}{v} = y_t\left(1 - \frac{\alpha_s}{4\pi}\frac{\delta_{m_t}}{\epsilon} + \mathcal{O}(\alpha_s^2) \right) .
\end{equation}
Finally, the wave functions renormalization of external particles is realized by simply multiplying the scattering amplitude by $\displaystyle \sqrt{Z_t}$ for each top quark and $\displaystyle \sqrt{Z_g}$ for each gluon. In the on-shell scheme these factors read~\cite{Melnikov:2000zc} 
\begin{align}
    Z_t &=1 - \frac{\alpha_s}{4\pi} \frac{\delta_t}{\epsilon} + \mathcal{O}(\alpha_s^2), \quad \text{with} \quad \delta_t = \delta_{m_t}, \\
    Z_g &=1 - \frac{\alpha_s}{4 \pi} \frac{\delta_g}{\epsilon} + \mathcal{O}(\alpha_s^2), \quad \text{with} \quad \delta_g = \frac{2}{3} N_h \,.
\end{align}

Combining everything together, the tree-level and one-loop contributions to the UV renormalized scattering amplitude read 
\begin{align}
    \mathcal{A}^{(0)}_{\rm r} &= \mathcal{A}^{(0)}, \notag\\
    \mathcal{A}^{(1)}_{\rm r} &= \mathcal{A}^{(1)} - \frac{\mathcal{A}^{(1)}_{\rm ct}}{\epsilon}, \quad \text{ with } \quad \mathcal{A}^{(1)}_{\rm ct} = (\beta_0 + \delta_{m_t} + \delta_t + \delta_g)\mathcal{A}^{(0)} + \delta_{m_t}\, \mathcal{A}^{(0)}_{m,\rm ct}\,,
    \label{eq:UV_Renormalized_amplitude}
\end{align}
where the subscript $\rm r$ refers to renormalized quantities and $\mathcal{A}^{(0)}_{m,\rm ct}$ is obtained applying the shift in Eq.~\eqref{eq:Mass_Counterterm} to the tree-level scattering amplitude.

After UV renormalization, the amplitude contains residual $\epsilon$-poles of pure IR origin. 
Their general structure is fully predicted in terms of lower-loop results~\cite{Catani:1998bh,Catani:2002hc,Becher:2009cu,Becher:2009kw}, 
thus the agreement between our left-over poles and their universal behavior will serve as a powerful check of our calculation.
For our purposes, we follow the approach of~\cite{Catani:2002hc} and define the IR-finite one-loop amplitude
\begin{equation}
     \mathcal{A}^{(1)}_{\rm{fin}} = \mathcal{A}^{(1)}_{\rm r} - \boldsymbol{I}_1(\mu^2,\epsilon) \mathcal{A}^{(0)}_{\rm r}\,.
    \label{eq:Catani_Formula}
\end{equation}

The explicit form of the insertion operator $\boldsymbol{I}_1(\mu^2,\epsilon)$ can be found in~\cite{Catani:2002hc},
but we report it here for completeness
\begin{align}
    &\boldsymbol{I}_1(\mu^2,\epsilon)=-\frac{\alpha_{\mathrm{s}}}{4 \pi} \frac{(4 \pi)^\epsilon}{\Gamma(1-\epsilon)} 2\sum_{j=1}^4 \frac{1}{\boldsymbol{T}_j^2} \sum_{k=1, k \neq j}^4 \boldsymbol{T}_j \cdot \boldsymbol{T}_k \notag\\
    & \times\left[\boldsymbol{T}_j^2\left(\frac{\mu^2}{2 p_i p_j}\right)^\epsilon\mathcal{V}_j\left(s_{j k}, m_j, m_k ; \epsilon\right)+\Gamma_j +\gamma_j \ln \left(\frac{\mu^2}{2 p_i p_j}\right)+\gamma_j+K_j+\mathrm{O}(\epsilon)\right].
    \label{eq:Catani_Formula_2}
\end{align}
The operators $\boldsymbol{T}_k$ act on the elements of the color space $\ket{\mathcal{C}_i}$ defined in Eq.~\eqref{eq:def_color_basis}, see e.g.~\cite{Catani:1998bh}.
Their form depends on the flavour of the corresponding external parton $k$, which could be either a gluon, $k=g$, or a top (antitop) quark, $k=t$.
The constants in Eq.~\eqref{eq:Catani_Formula_2} read
\begin{equation}
    \gamma_q=\frac{3}{2} C_{F}, \quad \gamma_g=\frac{11}{6} C_{A}-\frac{1}{3}  N_f,
\end{equation}
\begin{align}
    \Gamma_t = C_F\left(\frac{1}{\epsilon} - \frac{1}{2}\ln{\frac{\mu^2}{m_t^2}} -2 \right), \quad 
    \Gamma_g = \frac{1}{\epsilon}\gamma_g + \frac{1}{3}\ln\frac{\mu^2}{m_t^2},
\end{align}
and
\begin{equation}
    K_q = \left(\frac{7}{2} - \frac{\pi^2}{6}\right)C_F, \quad K_g=\left(\frac{67}{18}-\frac{\pi^2}{6}\right) C_A -\frac{5}{9} N_f\,.
\end{equation}

Following \cite{Catani:2002hc}, we split the function $\mathcal{V}_j$ into a singular part $\mathcal{V}^{(\mathrm{S})}_j$
and a non-singular one $\mathcal{V}^{(\mathrm{NS})}_j$. 
Since we are interested only in the poles structure, we report here only the singular piece. 
The form of $\mathcal{V}^{(\mathrm{S})}_j$ depends on the masses of the pair of particles $(jk)$. In the various configurations they read 
\begin{align}
    \mathcal{V}^{(\mathrm{S})}\left(s_{j k}, 0,0 ; \epsilon\right) &=  \frac{1}{\epsilon^2}, \notag\\
    \mathcal{V}^{(\mathrm{S})}\left(s_{j k}, m_j,0 ; \epsilon\right) &=  \mathcal{V}^{(\mathrm{S})}\left(s_{j k}, 0, m_j ; \epsilon\right) \notag\\
    &=  \frac{1}{2 \epsilon^2}+\frac{1}{2 \epsilon} \ln \frac{m_j^2}{s_{j k} - m_j^2}-\frac{1}{4} \ln ^2 \frac{m_j^2}{s_{j k} - m_j^2}-\frac{\pi^2}{12}, \notag\\
    & -\frac{1}{2} \ln \frac{m_j^2}{s_{j k} - m_j^2 } \ln \frac{s_{j k} - m_j^2}{s_{j k}}-\frac{1}{2} \ln \frac{m_j^2}{s_{j k}} \ln \frac{s_{j k} - m_j^2}{s_{jk}}, \notag\\
    \mathcal{V}^{(\mathrm{S})}\left(s_{j k}, m_j, m_k; \epsilon\right)= & \frac{1}{v_{j k}}\bigg[\frac{1}{\epsilon} \ln \sqrt{\frac{1-v_{jk}}{1+v_{jk}}}-\frac{1}{4} \ln ^2 \rho_{jk}^2-\frac{1}{4} \ln ^2 \rho_{kj}^2-\frac{\pi^2}{6}\notag\\
    &  +\ln \sqrt{\frac{1-v_{jk}}{1+v_{jk}}} \ln \left(\frac{s_{j k}}{s_{j k} - m_j^2 - m_k^2}\right)\bigg] .
\end{align}
In the last line, we defined the relative velocity
\begin{equation}
    v_{ij}=\sqrt{1-\frac{4m_i^2m_j^2}{(s_{ij}-m_i^2-m_j^2)^2}},
\end{equation}
and the auxiliary quantity
\begin{equation}
    \rho_{j k} = \sqrt{\frac{1-v_{j k} + \frac{2m_j^2}{s_{jk} - m_j^2 - m_k^2}}{1+v_{j k} + \frac{2m_j^2}{s_{jk} - m_j^2 - m_k^2}}}.
\end{equation}

Let us conclude this section with a final remark regarding the interplay between the renormalization 
procedure and our choice of regularization scheme. 
The relevant anomalous dimensions in the IR-poles prediction and part of the UV counterterms, 
see e.g. Eq.~\eqref{eq:topmassct}, were derived in CDR, where external states are $D$ dimensional. 
Formally, this corresponds to a different regularization scheme with respect to ours. 
However, this difference is immaterial in our scheme. 
Indeed, assume that external states were $D$-dimensional. The tensors $\mathcal{T}_i, i = 1,\dots,8$ defined in Eq.~\eqref{eq:def_helicity_tensor_basis}, would still be a valid choice,
but they alone would be insufficient to span the whole $D$-dimensional space. 
Thus, we would have to account for additional tensors $\Tilde{\mathcal{T}}_i$. 
Without loss of generality, these additional tensors $\Tilde{\mathcal{T}}_i$ can be chosen to be orthogonal to the original $\mathcal{T}_i$~\cite{Peraro:2019cjj,Peraro:2020sfm} such that 
\begin{equation}
    \sum_{spin} \mathcal{T}^{\dagger}_i\cdot\Tilde{\mathcal{T}}_j = 0.
\end{equation}
Therefore, the form factors $\mathcal{F}_i$ in Eq.~\eqref{eq:linear_comb_hel_form_factors}  can be computed in CDR
completely independently from the form factors 
$\Tilde{\mathcal{F}}_i$ corresponding to the extra tensors $\Tilde{\mathcal{T}}$.
By adding this extra set of tensors, the scattering amplitude in CDR can be expressed as
\begin{equation}
    \mathcal{A}^{\text{CDR}} = \sum_{i=1}^8\mathcal{F}_i\mathcal{T}_i + \sum_{j=1}^N\Tilde{\mathcal{F}}_j\Tilde{\mathcal{T}}_j.
\end{equation}
Upon performing the UV-renormalization and subtracting IR poles, the corresponding 
results in CDR become
\begin{align}
    \mathcal{A}^{\text{CDR}}_{\rm r} &= \sum_{i=1}^8\mathcal{F}_{i,\rm r}\mathcal{T}_i + \sum_{j=1}^N\Tilde{\mathcal{F}}_{j,\rm r}\Tilde{\mathcal{T}}_j, \\
    \mathcal{A}^{\text{CDR}}_{\rm fin} &= \sum_{i=1}^8\mathcal{F}_{i,\rm fin}\mathcal{T}_i + \sum_{j=1}^N\Tilde{\mathcal{F}}_{j,\rm fin}\Tilde{\mathcal{T}}_j.
\end{align}
The poles cancellation is realized at the level of individual form factors. This means
that one can perform UV renormalization on each individual form factor in CDR and,
if one neglects the extra $\Tilde{\mathcal{F}}_i$, one obtains the UV renormalized
amplitude in tHV.
Secondly, if one subtracts also all IR poles and computes the finite remainder,
since the extra tensors $\Tilde{\mathcal{T}}_j$ are zero for four-dimensional external states, the difference in 
the regularization schemes has no impact and the finite remainder in tHV is identical to the one in CDR.
\section{Amplitude and Master Integrals Calculation}
\label{sec:details_of_computation}

We compute the form factors in Eq.~\eqref{eq:helicity_formfactors_even_odd} by applying suitable combinations
of the projector operators defined in Eq.~\eqref{eq:def_projector} directly on the Feynman diagrams.
In particular, we generate all tree-level and one-loop Feynman diagrams with \texttt{QGRAF}~\cite{Nogueira:1991ex} 
and use \texttt{FORM}~\cite{Ruijl:2017dtg} to derive and apply the projectors on the individual Feynman diagrams. 
This requires performing color and Dirac algebra.
In this way, we express the helicity form factors in terms of linear combinations of scalar Feynman integrals.
The latter can be organized in four independent integral families and crossings thereof.
In contrast to~\cite{Chen:2022nxt}, where the the four integral families were studied individually, 
we consider all of them at once
in order to obtain a non-redundant integral basis for the complete scattering amplitude. 
This will also allow us
to uncover extra relations among the master integrals, which are not straightforwardly identified
by automated reduction programs.

We use the four independent integral families $Y={A,B,C,D}$~\cite{Chen:2022nxt}
\begin{equation}
    \label{eq:Def_Feynman_Integrals}
    I^Y_{\nu_1\nu_2\nu_3\nu_4\nu_5}(D) =  \int\frac{\text{d}^Dk}{i\pi^{\frac{D}{2}}}
    \frac{1}{P_1^{\nu_1}P_2^{\nu_2}P_3^{\nu_3}P_4^{\nu_4}P_5^{\nu_5}}, \quad Y\in\{A,B,C,D\}\,,
\end{equation}
where the inverse propagators take the form 
\begin{align} \label{eq:defPi}
    P_i = (k+r_i)^2-m_i^2.
\end{align}

\begin{figure}[h!]
        \centering
        \begin{subfigure}[b]{0.45\textwidth}
            \centering
            \includegraphics[width=0.8\textwidth]{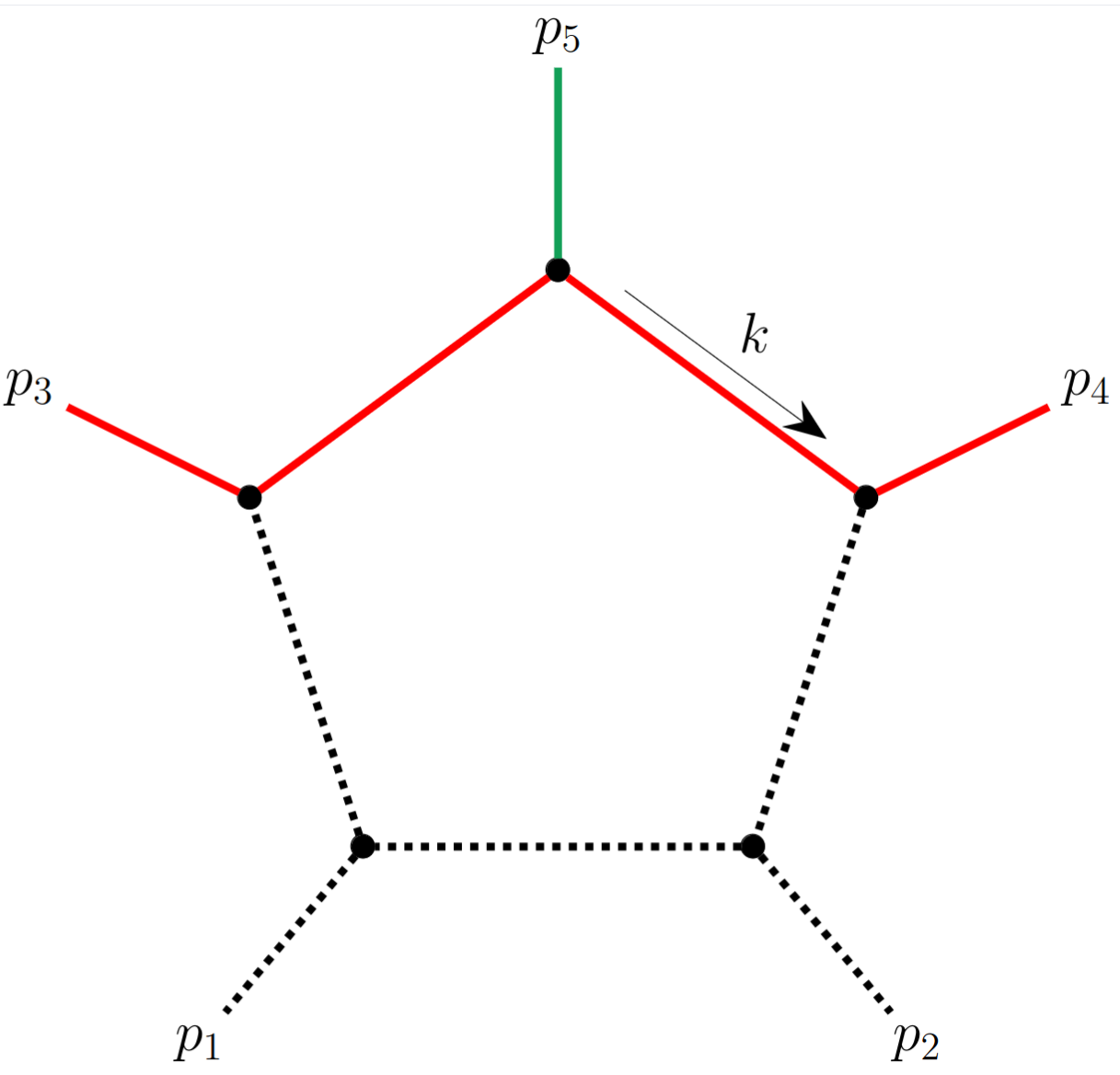}
            \caption[Topology A]%
            {{\small Topology $A$}}    
            \label{fig:Pentagon_Topo_A}
        \end{subfigure}
        %\hfill
        \begin{subfigure}[b]{0.45\textwidth}  
            \centering 
            \includegraphics[width=0.8\textwidth]{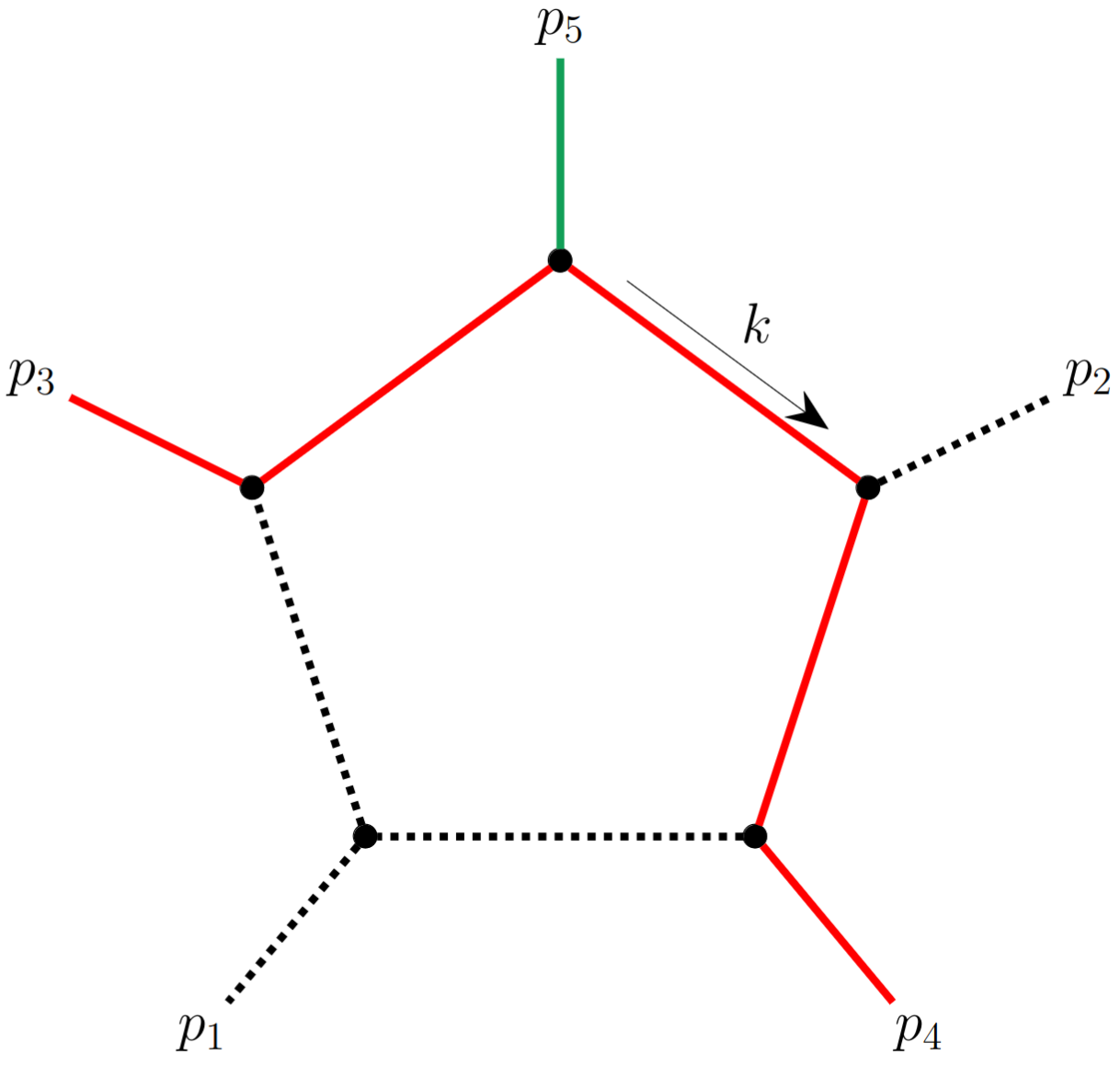}
            \caption[Topology B]%
            {{\small Topology $B$}}    
            \label{fig:Pentagon_Topo_B}
        \end{subfigure}
        \vskip\baselineskip
        \begin{subfigure}[b]{0.45\textwidth}   
            \centering 
            \includegraphics[width=0.8\textwidth]{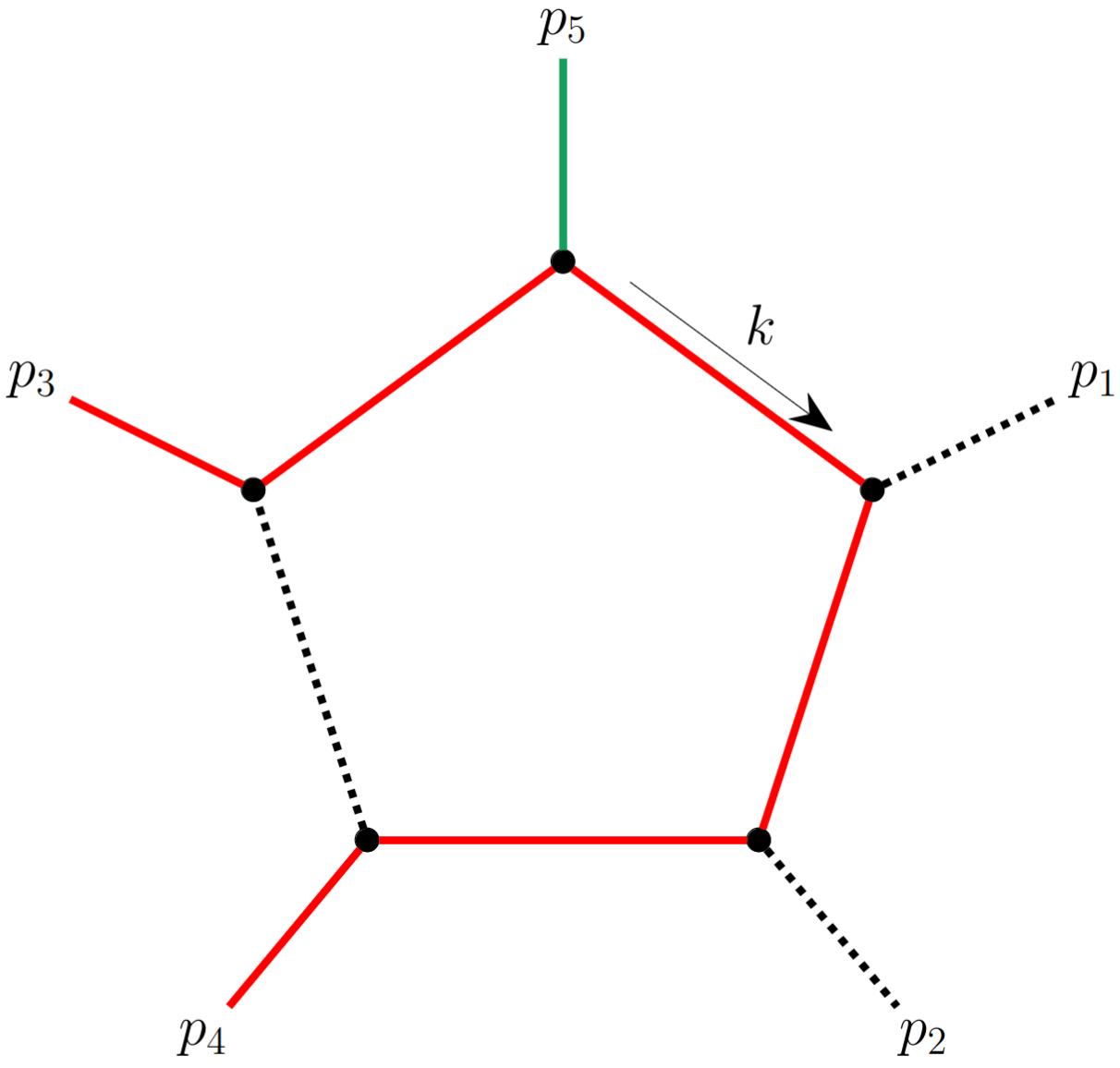}
            \caption[Topology C]%
            {{\small Topology $C$}}    
            \label{fig:Pentagon_Topo_C}
        \end{subfigure}
        %\hfill
        \begin{subfigure}[b]{0.45\textwidth}   
            \centering 
            \includegraphics[width=0.8\textwidth]{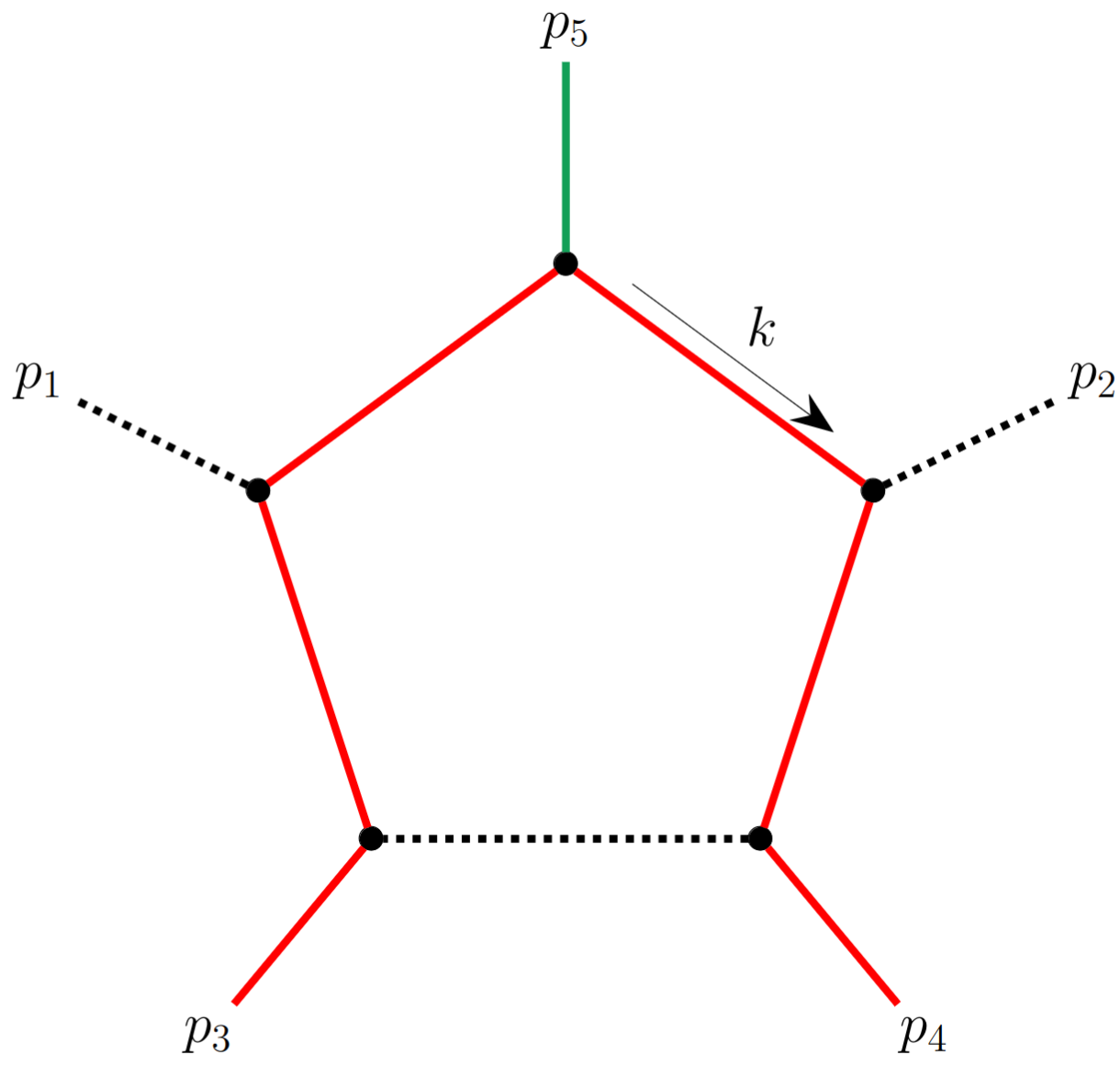}
            \caption[Topology D]%
            {{\small Topology $D$}}    
            \label{fig:Pentagon_Topo_D}
        \end{subfigure}
        \caption{\small Integral families for \tth\!\!. 
        Red lines denote massive propagators and external legs of mass $m_t$, 
        green lines denote massive propagators of mass $m_h$, and dotted lines are massless.} 
        \label{fig:Pentagon_Topos}
    \end{figure}

\begin{table}[th]
\centering
\begin{tabular}{|c|c|c|c|c|}
\hline
      & $A$                   & $B$                    & $C$ & $D$ \\ \hline
$P_1$ & $k^2 -m_t^2$          & $k^2 -m_t^2$           & $k^2 -m_t^2$               & $k^2 -m_t^2$   \\ 
$P_2$ & $(k + p_4)^2$         & $(k + p_2)^2 -m_t^2$   & $(k + p_1)^2 -m_t^2$       & $(k + p_2)^2 -m_t^2$ \\ 
$P_3$ & $(k +p_2 + p_4)^2$    & $(k +p_2 + p_4)^2$     & $(k +p_1 + p_2)^2 -m_t^2$  & $(k +p_2 + p_4)^2$  \\ 
$P_4$ & $(k -p_3-p_5)^2$      & $(k -p_3-p_5)^2$       & $(k -p_3-p_5)^2$           & $(k -p_1-p_5)^2 -m_t^2$  \\ 
$P_5$ &  $(k -p_5)^2 - m_t^2$ & $(k - p_5)^2 - m_t^2$  & $(k -p_5)^2 - m_t^2$       & $(k -p_5)^2 - m_t^2$  \\ \hline
\end{tabular}
\caption{Definition of the four independent integral families contributing at one loop.}
\label{tab:integralfamilies}
\end{table}
The definition of the four independent integral families is reported in Tab.~\ref{tab:integralfamilies},
see also Fig.~\ref{fig:Pentagon_Topos} for a graphical representation.
In addition, for each family $Y \in\{A,B,C,D\}$, we define the corresponding crossed families 
\begin{align}
    Y_{\text{x12}} \equiv Y_{\{p_1 \leftrightarrow p_2\}}, \;\; 
    Y_{\text{x34}} \equiv Y_{\{p_3 \leftrightarrow p_4\}},  \;\;
    Y_{\text{x12x34}} \equiv Y_{\{p_1 \leftrightarrow p_2, p_3 \leftrightarrow p_4\}}.
\end{align}

We reduce all the ensuing integrals using the IBP reduction programs \texttt{Reduze}~\cite{Studerus:2009ye} and \texttt{KIRA}~\cite{Maierhofer:2017gsa,Klappert:2020nbg}. 
We find $87$ apparently independent master integrals. 
However, both reduction codes miss five extra identities, reducing the number
of independent integrals to $82$. 
We group these relations into two \textit{box-symmetry relations},
\begin{align}
    I^{\text{B}}_{1 1 1 1 0}(D) &= I^{\text{Bx12}}_{1 1 1 1 0}(D)\,, \qquad
    I^{\text{Bx34}}_{1 1 1 1 0}(D) = I^{\text{Bx12x34}}_{1 1 1 1 0}(D)\,,
    \label{eq:symmetry_relations_box}
\end{align}
and three  \textit{triangle-symmetry relations}

\begin{align}
  0 &= (1 - s_{12} - s_{24}) I^{\text{B}}_{1 1 0 1 0}(D)    - (1 - s_{24}) I^{\text{B}}_{1 1 1 0 0}(D) \notag\\
    &- (1 - s_{12} - s_{14}) I^{\text{Bx12}}_{1 1 0 1 0}(D) + (1 - s_{14}) I^{\text{Bx12}}_{1 1 1 0 0}(D), \notag\\[6pt]
  0 &= (1 - s_{12} - s_{13}) I^{\text{Bx12x34}}_{1 1 0 1 0}(D) - (1 - s_{13}) I^{\text{Bx12x34}}_{1 1 1 0 0}(D) \notag\\
    &- (1 - s_{12} - s_{23}) I^{\text{Bx34}}_{1 1 0 1 0}(D)    + (1 - s_{23}) I^{\text{Bx34}}_{1 1 1 0 0}(D), \\[6pt]
   0 &= ( s_{12} + s_{23} + s_{24}-2) I^{\text{B}}_{1 1 0 0 1}(D) - ( s_{12} + s_{13} + s_{14}-2) I^{\text{Bx12}}_{1 1 0 01}(D) \notag\\
     &+ (2 - s_{23} - s_{24}) I^{\text{C}}_{0 1 1 0 1}(D) - (2 - s_{13} - s_{14}) I^{\text{Cx12}}_{0 1 1 0 1}(D). \notag
    \label{eq:Triangle_relations}
\end{align}

We discovered these extra relations as follows. 
Following~\cite{Chen:2022nxt}, we started by introducing a canonical basis  of 
master integrals. 
We then evaluated the canonical integrals numerically to high precision, roughly 300 significant digits, up to transcendental weight six, 
using our refined implementation of the \texttt{AMFlow} algorithm~\cite{Liu:2022chg}, see Sec.~\ref{sec:seminumerical}.
Using the \texttt{PSLQ} algorithm~\cite{Ferguson:PSLQ}, we searched for linear relations among the various epsilon coefficients
of the canonical master integrals.
As expected, at low transcendental weights one finds many such relations, 
since not all analytic structures contribute to the poles and to the finite part of the scattering amplitude. 
Surprisingly, we observed that the five relations in Eq.~\eqref{eq:symmetry_relations_box} and Eq.~\eqref{eq:Triangle_relations} 
hold identically to all the computed $\epsilon$ orders.  
This provided us with the motivation to prove 
them analytically to all orders.

Let us start with the box-symmetry relations Eq.~\eqref{eq:symmetry_relations_box}. 
We prove them in Feynman parameter representation. 
The graph polynomials for  $I^{\text{B}}_{1 1 1 1 0}(D)$ and $I^{\text{Bx12}}_{1 1 1 1 0}(D)$ are~\cite{Bogner:2010kv}
\begin{align}
    \mathcal{U}_{11110}  &= x_1 + x_2 + x_3 +x_4,  \\
    \mathcal{F}^{\text{B}}_{11110}  &= -s_{12} x_1x_4 - s_{14} (x_1 + x_2)x_4 - s_{24} x_1 (x_3 + x_4) \notag\\&
    + m_t^2 \left[x_1^2 + x_2 (x_2 + x_4) + x_1 (2 x_2 + x_3 + 2 x_4)\right], \notag\\
    \mathcal{F}^{\text{Bx12}}_{11110}  &= -s_{12} x_1x_4 - s_{13} (x_1 + x_2)x_4 - s_{23} x_1 (x_3 + x_4) \notag\\
    &+ m_t^2 \left[x_1^2 + x_2 (x_2 + x_4) + x_1 (2 x_2 + x_3 + 2 x_4)\right].
    \label{eq:graph_polynomials}
\end{align}
Under the variable change 
\begin{equation}
    (x_1, x_2, x_3, x_4)  \to \left(\frac{(x_1 + x_2) x_4}{x_3 + x_4}, \frac{(x_1 + x_2) x_3}{x_3 + x_4},
    \frac{x_2 (x_3 + x_4)}{x_1 + x_2},  \frac{x_1 (x_3 + x_4)}{x_1 + x_2}\right),
\end{equation}
$\mathcal{F}^{\text{Bx12}}_{11110}$ maps to $\mathcal{F}^{\text{B}}_{11110}$. The Jacobian of this transformation is $1$ and the 
$\delta$-function constraint in the integrand is mapped onto itself. Hence, the equivalence is proved.
The second box symmetry relation then follows from crossing the momenta $p_3\leftrightarrow p_4$ in Eq.~\eqref{eq:graph_polynomials}.

To prove the triangle-symmetry relations Eq.~\eqref{eq:Triangle_relations} instead, we start from their differential equations.
We first computed their derivatives 
with respect to all variables and verified that the derivatives add to zero. 
As a consequence, each combination can at most be equal to a constant. 
By taking a suitable limit, we show that this constant is zero in all triangle-symmetry relations.
In the first relation, the second graph polynomials are 
\begin{align}
\label{eq:graph_polynomials_triangle_1}
%B11100
%    \mathcal{U}_{11100} &= \\
    \mathcal{F}^{\text{B}}_{11100} &= - x_2x_3m_t^2 - x_1x_3s_{24} + (x_1+x_2)m_t^2 + (x_1+x_2)(x_1 + x_2 + x_3 )m_t^2, \notag\\
    \mathcal{F}^{\text{Bx12}}_{11100} &= - x_2x_3m_t^2 - x_1x_3s_{14} + (x_1+x_2)m_t^2 + (x_1+x_2)(x_1 + x_2 + x_3 )m_t^2, \\[7pt]
    %B11010
  %  \mathcal{U}_{11010} &= x_1 + x_2 + x_4 \\
    \mathcal{F}^{\text{B}}_{11010} &= x_1x_4(m_t^2 - s_{12} - s_{14}) - x_2x_4s_{24} + (x_1+x_2)m_t^2 + (x_1+x_2)(x_1 + x_2 + x_4 )m_t^2, \notag\\
    \mathcal{F}^{\text{Bx12}}_{11010} &= x_1x_4(m_t^2 - s_{12} - s_{24}) - x_2x_4s_{14} + (x_1+x_2)m_t^2 + (x_1+x_2)(x_1 + x_2 + x_4 )m_t^2. \notag
\end{align}
From Eq.~\eqref{eq:graph_polynomials_triangle_1}, it is clear that in the limit $s_{24} \to s_{14}$ 
\begin{equation}
    I^{\text{B}}_{11100} = I^{\text{Bx12}}_{11100} \quad \text{ and } \quad I^{\text{B}}_{11010} = I^{\text{Bx12}}_{11010}.
\end{equation}
This limit is smooth, so the terms cancel pairwise. This completes the proof of the first triangle-symmetry relation Eq.~\eqref{eq:Triangle_relations}.
The second triangle-symmetry relation directly follows from the first relation upon permuting $p_3\leftrightarrow p_4$. 

In the last relation, the second graph polynomials are 
\begin{align}
    \mathcal{F}^{\text{B}}_{11001} &= x_1x_5(4m_t^2 - s_{12} - s_{13} - s_{14} - s_{23}- s_{24}-s_{34}) +x_2x_5(2m_t^2 - s_{13} - s_{14} - s_{34})  \notag\\ & + (x_1+x_2+x_5)^2 m_t^2, \notag\\ 
    \mathcal{F}^{\text{Bx12}}_{11001} &= x_1x_5(4m_t^2 - s_{12} - s_{13} - s_{14} - s_{23}- s_{24}-s_{34}) +x_2x_5(2m_t^2 - s_{23} - s_{24} - s_{34})  \notag\\ & + (x_1+x_2+x_5)^2 m_t^2, \notag\\[7pt]
    \mathcal{F}^{\text{C}}_{01101}    &= x_2x_5(2m_t^2 - s_{23} - s_{24} - s_{34}) - x_3x_5s_{34} + (x_2+x_3+x_5)^2 m_t^2, \notag\\
    \mathcal{F}^{\text{Cx12}}_{01101} &= x_2x_5(2m_t^2 - s_{13} - s_{14} - s_{34}) - x_3x_5s_{34} + (x_2+x_3+x_5)^2 m_t^2. 
    \label{eq:graph_polynomials_triangle_2}
\end{align}
From Eq.~\eqref{eq:graph_polynomials_triangle_2}, it is clear that in the limit $s_{23}\to s_{13}, s_{24}\to s_{14}$ 
\begin{equation}
    I^{\text{B}}_{11001} = I^{\text{Bx12}}_{11001} \quad \text{ and } \quad I^{\text{C}}_{01101} = I^{\text{Cx12}}_{01101}.
\end{equation}
Similarly to the first triangle-symmetry relation, the terms cancel pairwise in this limit, thus, the constant is zero. 

We stress here that, somewhat unexpectedly, the differential equations for the redundant set of $87$ master integrals 
satisfy the integrability condition, which is sometimes used as a check of the fact that all 
relations among master integrals have been identified, see for example~\cite{Adams:2018bsn}.

In conclusion, we choose the following $82$ master integrals
\begin{alignat}{4}
%    \begin{array}{lllll}
 &   I_{1}=I^{\text{A}}_{00102}(D),     && \quad I_{2}=I^{\text{A}}_{01002}(D),     && \quad I_{3}=I^{\text{A}}_{01011}(D),    && \quad I_{4}=I^{\text{A}}_{01111}(D), \notag \\
 &   I_{5}=I^{\text{A}}_{02010}(D),     && \quad I_{6}=I^{\text{A}}_{10011}(D),     && \quad I_{7}=I^{\text{A}}_{10101}(D),    && \quad I_{8}=I^{\text{A}}_{10111}(D), \notag \\
 &   I_{9}=I^{\text{A}}_{11001}(D),     && \quad I_{10}=I^{\text{A}}_{11010}(D),    && \quad I_{11}=I^{\text{A}}_{11011}(D),   && \quad I_{12}=I^{\text{A}}_{11101}(D),\notag  \\
 &   I_{13}=I^{\text{A}}_{11110}(D),    && \quad I_{14}=I^{\text{A}}_{11111}(D),    && \quad I_{15}=I^{\text{A}}_{20000}(D),   && \quad I_{16}=I^{\text{A}}_{20001}(D),\notag  \\
 &   I_{17}=I^{\text{A}}_{20010}(D),    && \quad I_{18}=I^{\text{A}}_{20100}(D),    && \quad I_{19}=I^{\text{Ax12}}_{00102}(D),&& \quad I_{20}=I^{\text{Ax12}}_{01111}(D),\notag  \\
 &   I_{21}=I^{\text{Ax12}}_{10101}(D), && \quad I_{22}=I^{\text{Ax12}}_{10111}(D), && \quad I_{23}=I^{\text{Ax12}}_{11101}(D),&& \quad I_{24}=I^{\text{Ax12}}_{11110}(D), \notag \\
 &   I_{25}=I^{\text{Ax12}}_{11111}(D), && \quad I_{26}=I^{\text{Ax12}}_{20100}(D), && \quad I_{27}=I^{\text{B}}_{01011}(D),   && \quad I_{28}=I^{\text{B}}_{01101}(D), \notag \\
 &   I_{29}=I^{\text{B}}_{01111}(D),    && \quad I_{30}=I^{\text{B}}_{02001}(D),    && \quad I_{31}=I^{\text{B}}_{11001}(D),   && \quad I_{32}=I^{\text{B}}_{11010}(D), \notag  \\
 &    I_{33}=I^{\text{B}}_{11011}(D),      && \quad I_{34}=I^{\text{B}}_{11100}(D),       && \quad I_{35}=I^{\text{B}}_{11101}(D),       && \quad I_{36}=I^{\text{B}}_{11110}(D),   \notag      \\
 &   I_{37}=I^{\text{B}}_{11111}(D),       && \quad I_{38}=I^{\text{Bx12}}_{01011}(D),    && \quad I_{39}=I^{\text{Bx12}}_{01101}(D),    && \quad I_{40}=I^{\text{Bx12}}_{01111}(D), 
 \label{eq:Master_indegrals_d4} \\
 &   I_{41}=I^{\text{Bx12}}_{02001}(D),    && \quad I_{42}=I^{\text{Bx12}}_{11001}(D),    && \quad I_{43}=I^{\text{Bx12}}_{11011}(D),    && \quad I_{44}=I^{\text{Bx12}}_{11100}(D), \notag    \\ 
 &   I_{45}=I^{\text{Bx12}}_{11101}(D),    && \quad I_{46}=I^{\text{Bx12}}_{11111}(D),    && \quad I_{47}=I^{\text{Bx12x34}}_{11010}(D), && \quad I_{48}=I^{\text{Bx12x34}}_{11011}(D),\notag  \\ 
 &   I_{49}=I^{\text{Bx12x34}}_{11101}(D), && \quad I_{50}=I^{\text{Bx12x34}}_{11111}(D), && \quad I_{51}=I^{\text{Bx34}}_{11010}(D)   , && \quad I_{52}=I^{\text{Bx34}}_{11011}(D),  \notag   \\
 &   I_{53}=I^{\text{Bx34}}_{11100}(D),    && \quad I_{54}=I^{\text{Bx34}}_{11101}(D),    && \quad I_{55}=I^{\text{Bx34}}_{11110}(D)   , && \quad I_{56}=I^{\text{Bx34}}_{11111}(D), \notag     \\
 &   I_{57}=I^{\text{C}}_{00201}(D),       && \quad I_{58}=I^{\text{C}}_{01101}(D),       && \quad I_{59}=I^{\text{C}}_{01111}(D)      , && \quad I_{60}=I^{\text{C}}_{10101}(D),    \notag    \\
 &   I_{61}=I^{\text{C}}_{10110}(D),       && \quad I_{62}=I^{\text{C}}_{10111}(D),       && \quad I_{63}=I^{\text{C}}_{11100}(D)      , && \quad I_{64}=I^{\text{C}}_{11101}(D),    \notag     \\ 
 &   I_{65}=I^{\text{C}}_{11110}(D),       && \quad I_{66}=I^{\text{C}}_{11111}(D),       && \quad I_{67}=I^{\text{C}}_{20100}(D)      , && \quad I_{68}=I^{\text{Cx12}}_{01111}(D), \notag    \\
 &   I_{69}=I^{\text{Cx12}}_{11101}(D),    && \quad I_{70}=I^{\text{Cx12}}_{11110}(D),    && \quad I_{71}=I^{\text{Cx12}}_{11111}(D)   , && \quad I_{72}=I^{\text{Cx12x34}}_{01111}(D), \notag \\
 &   I_{73}=I^{\text{Cx12x34}}_{11110}(D), && \quad I_{74}=I^{\text{Cx12x34}}_{11111}(D), && \quad I_{75}=I^{\text{Cx34}}_{01111}(D)   , && \quad I_{76}=I^{\text{Cx34}}_{10110}(D), \notag    \\
 &   I_{77}=I^{\text{Cx34}}_{10111}(D),    && \quad I_{78}=I^{\text{Cx34}}_{11110}(D),    && \quad I_{79}=I^{\text{Cx34}}_{11111}(D)   , && \quad I_{80}=I^{\text{D}}_{11011}(D),    \notag    \\
 &   I_{81}=I^{\text{D}}_{11111}(D),       && \quad I_{82}=I^{\text{Dx12}}_{11111}(D).    &&                                      &&  \notag
%   \end{array}
\end{alignat}

\section{Challenges of the Analytic Calculation}
\label{sec:analaytic_challenges}
A full analytic calculation of the scattering amplitudes requires two main ingredients. First, a full reduction to master integrals and, second, 
analytic solutions for the master integrals in terms of independent iterated integrals.
Despite this being only a one-loop calculation, the complexity of the kinematics, 
which is a consequence of the large
number of scales the problem depends on, renders both steps very challenging. In the following,
we elaborate on the two problems separately. 

Let us start with the reduction to master integrals.
Contrary to the typical situation at two loops and higher, the relevant tables of IBP identities 
can be  obtained easily in compact form using automated codes as \texttt{Reduze} and \texttt{KIRA}. 
On the other hand, their insertion into the unreduced amplitude, either using standard programs as \texttt{Fermat}~\cite{lewis_fermat_2008} and \texttt{FORM}, or more specialized tools as \texttt{FiniteFlow}~\cite{Peraro:2019cjj} and \texttt{Firefly}~\cite{Klappert:2020nbg},
requires a lot of care.
In fact, due to the large number of scales involved, 
the overall size of the resulting rational functions increases considerably, 
which renders both their symbolic manipulation 
and their numerical evaluations non-trivial.
Different approaches can be attempted to simplify these rational functions. 
These include changing bases of master integrals (using six-dimensional pentagons and boxes),
expanding the amplitudes in $\epsilon$ in terms of independent transcendental functions, and 
using different versions of multivariate partial fraction decomposition. 
In particular, the latter has the potential 
to produce compact final results, but all available tools cannot easily derive
the partial fraction identities required up to order $\mathcal{O}(\epsilon^2)$, as we will comment on below.

More explicitly, in order to cope with intermediate expression swell originating from 
multivariate rational functions,
analytic reconstruction based on finite-field methods~\cite{vonManteuffel:2014ixa,Peraro:2019svx,Klappert:2020nbg}
is often the preferred computational strategy.
The basic idea is to extract an algebraic expression from numerical samples. This makes sense especially 
if the final result
is expected to be substantially simpler than the intermediate stages of the calculation, 
since the number of required samples depends on the polynomial degree and the number of variables. 
More precisely, it scales exponentially with the polynomial degree 
and it increases by an order of magnitude for each additional variable.
In our specific case we have to manipulate polynomials in six variables of degree up to $\mathcal{O}(50)$. 
Despite applying various improvements like denominator matching~\cite{vonManteuffel:2014ixa,Heller:2021qkz}, 
massive spinor-helicity formalism, 
reconstructing partial-fraction decomposed results~\cite{Heller:2021qkz,DGPS}, or improving the 
time-per-sample evaluation~\cite{Magerya:2022hvj}, generating enough sample points to reconstruct the
final rational functions remains prohibitive.
In contrast, and contrary to naive expectations, we found that the most efficient way to manipulate
these complicated rational functions is using \texttt{Fermat}.
We suspect that the main reason is that \texttt{Fermat} relies on Zippel's 
algorithm~\cite{ZIPPEL1990375} for the numerical reconstruction in the \texttt{GCD} algorithm, 
which is expected to scale better starting from rational functions that depend on six or more different scales.

In this context, some comments about the choice of basis of master integrals are in order, as they directly
impact the complexity of the rational functions involved.
For this discussion, it is convenient to decompose the partial amplitudes in Eq.~\eqref{eq:colorF} 
in powers of $N_c$, $N_f$ and $N_h$,
\begin{equation*}
  \mathcal{A}^{(1)}_{i} = N_c \left(\mathcal{A}^{(1,1)}_{i} + \frac{N_f+N_h}{N_c} \mathcal{A}^{(1,0)}_{i} + \frac{1}{N^2_c} \mathcal{A}^{(1,-1)}_{i}\right),
  \;\text{with}\;i=1,2\quad\text{and}\quad
    \mathcal{A}^{(1)}_3 = A^{(1,0)}_3,
\end{equation*}
where $N_f$ is the number of light quarks and $N_h$ the number of heavy ones.
In the standard master integral basis of Eq.~\eqref{eq:Master_indegrals_d4}, the Tadpole coefficients in $\mathcal{A}^{(1,1)}_{i}$ 
dominate the scattering amplitude's complexity. 
The largest single coefficient is around $650$ MB large. 
Interestingly, we find that rotating to a canonical basis (which, modulo irrelevant prefactors, means using dotted bubbles and six-dimensional pentagons) 
shifts part of the complexity from $\mathcal{A}^{(1,1)}_{i}$ to $\mathcal{A}^{(1,-1)}_{i}$ and $\mathcal{A}^{(1,0)}_3$, 
such that ultimately the overall size of the symbolic expressions (in \texttt{GCD} form) increases. 
This increase in complexity originates from  denominators of polynomial degree 3 and 4 which 
were previously implicit in the integral definitions, and become explicit, once one
writes the four-dimensional pentagons in terms of their six-dimensional counterparts.
Nevertheless, opting for a canonical basis partly simplifies the structure of the helicity form factors,
meaning that the spurious Gram determinant $1/\Delta$ 
cancels in the integral coefficients of the tadpole, bubbles, triangles, and most boxes.

Moving from this observation, we attempted to decompose the coefficients of the canonical
master integrals in partial fractions, using \texttt{MultivariateApart}~\cite{Heller:2021qkz} and \texttt{Singular}~\cite{DGPS}, 
and found, for some of them, impressive simplifications up to a factor $\mathcal{O}(500)$.
However, we did not manage to perform a full decomposition for those coefficients involving 
degree 3 or 4 polynomials, since these public programs
are not able to complete a Groebner basis calculation for the denominator factors involved.
This preliminary findings nevertheless indicate that it would be extremely interesting to experiment
with partial fraction decomposition further, in particular employing newly developed ideas to reconstruct
rational functions directly in partial fractioned form~\cite{Chawdhry:2023yyx}.
\newline

Let us now consider the analytic calculation of the master integrals.
A canonical basis for a five-point amplitude is readily derived in terms of 
two-dimensional tadpoles and bubbles and six-dimensional pentagons.
The corresponding system of differential equations can be obtained algorithmically,
but its analytic solution in terms of iterated integrals is not straightforward.
In fact, the knowledge of the alphabet allows, at least in principle, for the expansion of the canonical master integrals 
in terms of linearly-independent iterated integrals. Additional simplifications in the scattering amplitudes 
could then be made manifest if these explicit expressions are used. 
In practice, however, the alphabet contains a large number of letters, which themselves include many 
different square roots. This obscures analytic relations between the iterated integrals.  
In particular, it is worth mentioning that naively one finds letters with occurrences of various double square roots. 
While we were able to eliminate all of them by suitable recombination of the relevant letters,
the analytic properties of the resulting integrals remain involved. 
As an example, consider the two double square roots 
\begin{align}
    r_{\pm} &= \sqrt{ 2 s_{14} - s_{13}^2 - 2 s_{13} (s_{34} - 4)  - (s_{34} - 8) s_{34} - 17 \pm (s_{13} + s_{34} - 5)r},
    \label{eq:double_roots_original}
\end{align}
with 
\begin{equation}
    r = \sqrt{s_{13}^2+2 s_{13} (s_{34}-3)-4 s_{14}+(s_{34}-3)^2}.
\end{equation}
Individually, it is not possible to remove the double square roots, but in the product, 
sum and difference, these square roots drop. The relevant identities necessary
depend on the prescription one uses for the analytic continuation of the roots themselves in different
regions of the phase space. 
For example for the product one easily finds
\begin{equation}
    r_+ \cdot r_- = \pm2 (s_{13}-s_{14}+s_{34}-4)\,.
    \label{eq:product_doubleroots}
\end{equation}
We notice that the sign in Eq.~\eqref{eq:product_doubleroots} is phase-space point dependent.
Depending on that sign, the sum and difference of the double square roots become 
\begin{equation}
    \begin{array}{c|cc}
      \quad  & \mathbf{+} & \mathbf{-}  \\
      \hline
        r_+ + r_-\quad &\quad - \sqrt{-2}  (s_{13}+s_{34}-5) &   \sqrt{-2} r\\
        r_+ - r_-\quad &\quad \sqrt{-2} r & - \sqrt{-2}  (s_{13}+s_{34}-5)\,.
    \end{array}
\end{equation}
We can use these type of relations to remove all double roots.  In doing that, one has to pay extreme attention
to the branch cut structure of all involved roots, in order to avoid inconsistent manipulations. 
It is worth recalling here that there have proposals been to address the issue of non-rationalizable 
square roots, 
for example through algorithms that allow to integrate the (canonical) differential equations directly in terms of polylogarithms, 
see e.g.~\cite{Heller:2021qkz,Kreer:2021sdt}. Nevertheless, devising a general approach which works for complicated alphabets as the 
one encountered here, remains an outstanding problem.

As discussed above, the main reason why one would like to solve the master integrals analytically in terms of 
an independent set of transcendental functions, is to make all simplifications manifest in the corresponding rational
functions. This is particularly important when the higher $\epsilon$ orders of the one-loop amplitudes
are used to subtract IR poles of the two-loop ones and define the corresponding finite reminders.
Interestingly, if one has a canonical basis available, one can hope to obtain comparable
simplifications using an alternative and substantially simpler approach.
In particular, as we already observed, one can use the \texttt{PSLQ} algorithm to determine
relations among the various coefficients of the Laurent expansion in 
$\epsilon$ of the master integrals, see the discussion in Section~\ref{sec:details_of_computation}. 
These relations are a consequence of the fact that, in particular at lower orders in $\epsilon$, the number of 
independent transcendental functions is smaller than the number of master integrals in $D$ dimensions. 
In our specific case, in addition to five exact relations, we 
found $171$ relations valid for specific orders in the Laurent expansion.
By inserting them into the $\epsilon$-expanded helicity form factors, we could obtain extremely compact expressions 
for the amplitude poles, which as expected resemble the complexity of the corresponding tree-level expressions. 
We stress that these relations are obtained without any explicit analytic calculation of the master integrals. 
Unfortunately, starting from the finite remainder, the resulting expressions remain rather complicated and, in particular
when considering the higher orders in $\epsilon$, no substantially simplifications were observed.
This is expected, since the higher $\epsilon$ orders are spurious artifacts of dimensional regularization. 
On the other hand, we believe that it could be worth to investigate this approach as a tool
to remove spurious higher-orders in $\epsilon$ from the finite remainder of the corresponding two-loop amplitude.
In particular, if one has an $\epsilon$-factorized basis at disposal and if one can make sense of concepts
as transcendental weight and purity of the ensuing functions, one can use this approach to determine all relations among
the Laurent coefficients of the two-loop master integrals and the transcendental functions
which appear in the IR subtraction formulas. Using all these relations consistently, 
one can then attempt to write the resulting
finite remainders in terms of a minimal set of transcendental functions, making all simplifications manifest.

In summary, it is possible to derive the analytic integral coefficients and simplify them to some extent, 
but results remain cumbersome. For this reason, we decided to switch to a semi-numerical approach, as discussed in the following section, which outperforms the analytic one both in evaluation time and memory usage.
\section{Scattering Amplitude Evaluation in Auxiliary Mass Flow}
\label{sec:seminumerical}

In the previous section we discussed the challenges arising from symbolic manipulations of the analytic amplitude. 
Here, we discuss the semi-analytic approach we adopted to move from the unreduced amplitude to the final numerical result.
This is based on the insertion of numerical IBP identities and on the evaluation of the master integrals 
using our one-loop specific implementation of the Auxiliary Mass Flow (AMF) algorithm~\cite{Liu:2017jxz,Liu:2022chg}. 

We start by describing the improvements to the AMF method, which itself relies on two main steps. 
The first one is the introduction of an auxiliary mass parameter $\eta$ into some of the propagators
\begin{align}
    P_i \to P_i - \eta,
\end{align}
and the construction of the differential equations with respect to $\eta$, for the corresponding master integrals. 
The latter task is carried out in combination with public IBP solvers such as~\cite{Lee:2013mka,Smirnov:2019qkx,Peraro:2019svx,Klappert:2020nbg}.
The second one is the numerical solution of the differential equations, 
evolving the differential equation from $\eta=\infty$ to $\eta=\mathrm{i}0^-$ in order to recover the result in the physical region.

At one loop, it is possible to bypass the IBP reduction step which results in a significant improvement of the numerical integration 
performances. 
To achieve this, we need to introduce $\eta$ in \emph{all} propagators. Then, one can prove that
the following formula holds\footnote{This can be derived by simply combining Eq.~(14) and Eq.~(15) of \cite{Duplancic:2003tv}.}
\begin{align}\label{eq:oneloopode}
    (2 z_0\eta -C)\frac{\partial}{\partial\eta}I_{\nu_1\cdots\nu_K}(D) = (&D-1-N)z_0 I_{\nu_1\cdots\nu_K}(D)+\sum_{i=1}^K z_i I_{\nu_1\cdots\nu_i-1\cdots\nu_K}(D-2)
\end{align}
for $\nu_1,\ldots,\nu_K\geq 1$, where the coefficients $C, z_0, z_1, \ldots, z_K$ are defined through
\begin{align}
\left(
\begin{array}{cccc}
0 & 1 & \cdots & 1\\
1 & r_{11} & \cdots & r_{1K}\\
\vdots & \vdots & \ddots & \vdots\\
1 & r_{K1} & \cdots & r_{KK}
\end{array}
\right)\cdot\left(
\begin{array}{c}
-C\\
z_1\\
\vdots\\
z_K\end{array}\right)=
\left(
\begin{array}{c}
z_0\\
0\\
\vdots\\
0\end{array}\right),
\end{align}
with $r_{ij} = (r_i - r_j)^2 -m_i^2 - m_j^2$.
If $z_0\neq0$ or $C\neq0$, Eq.~\eqref{eq:oneloopode} represents a differential equation for $I_{\nu_1\cdots\nu_K}(D)$, whose inhomogeneous part is comprised of integrals in $(D-2)$ dimension with smaller sum of propagator powers. 

At some singular phase space points, $z_0$ and $C$  vanish simultaneously. In these cases, Eq.~\eqref{eq:oneloopode} is a linear relation among the integrals instead of a differential equation. Without loss of generality, let us assume $z_1\neq0$. After the substitution $D\to D+2$ and $\nu_1\to \nu_1+1$,  Eq.~\eqref{eq:oneloopode} reads
\begin{align}\label{eq:oneloopred}
    I_{\nu_1\cdots\nu_K}(D) = -\sum_{i=2}^K \frac{z_i}{z_1} I_{\nu_1+1\cdots\nu_i-1\cdots\nu_K}(D).
\end{align}
We obtain the differential equation for $I_{\nu_1\cdots\nu_K}(D)$ by differentiating Eq.~\eqref{eq:oneloopred} with respect to $\eta$ 
and applying Eq.~\eqref{eq:oneloopode} to the integrals on the right-hand side. 
Thus, for any integral, we derive an individual differential equation. Using Eq.~\eqref{eq:oneloopode} recursively, we construct a closed system of differential equations for the full basis of master integrals. We then solve it with the standard AMF method~\cite{Liu:2022chg,Liu:2017jxz}.

We further enhance the computation efficiency by using a numerical fit~\cite{Liu:2022chg}. In this approach, we insert small rational numbers for $\epsilon$ in the unreduced scattering amplitude, in the IBP relations and in the integrals. 
In order to recover the $\epsilon$-expanded scattering amplitude
\begin{align}
    \mathcal{A} = \frac{1}{\epsilon^2}\sum_{i=0}^{N} f_i\epsilon^i,
    \label{eq:eps_expanded_amplitude}
\end{align}
we fit the coefficients $f_i$ against the numerical evaluations.
If the numerical samples are extracted with  an high enough number of correct digits $p_0$, 
the approximate value of $f_i$, denoted with $\Bar{f}_i$, then has a relative accuracy~\cite{Liu:2022chg}
\begin{equation}
    \label{eq:precision_eps_fit} \delta_i \equiv \left|\frac{\Bar{f}_i - f_i}{f_i}\right| \sim\left(\frac{r}{R}\right)^{n - i}\,,\quad 0\leq i\leq n-1.
\end{equation}
Here, $r$ is the absolute magnitude of the $\epsilon$ values, $R$ is the convergence radius of the expansion~\cite{Liu:2022chg}
and $n$ is the number of different $\epsilon$ evaluations.   
Based on explicit tests, we found that if a number of correct digits $p$ is desired, the optimal settings are
\begin{align}
    n = 8, \quad \quad r = 10^{-(p/4+2)} \quad\quad\text{and}\quad \quad p_0 = 2p+20\, .
\end{align}

Since we avoid any symbolic manipulations on the analytic expressions, this significantly improves the overall computational efficiency. 
Furthermore, all substantial cancellations take place at the level of the numerical samples, rather than at the level of the $\epsilon$-expansions. Reaching the desired accuracy for the former is much more cost-effective than for the latter.

We have implemented the aforementioned methods in the \texttt{Mathematica} package \texttt{TTH}. Using this package, we are able to compute the tree one-loop interference
up to $\epsilon^2$ for any phase-space point within a few minutes.

\section{Results and checks}
\label{sec:results}

Our main result is the \texttt{Mathematica} package \texttt{TTH} which can be downloaded from git 
\begin{center}
    \GIT\,.
\end{center}
The package provides three functions: \texttt{TTHAmplitudeTreeTree}, \texttt{TTHAmplitudeLoopTree}, and \texttt{TTHUVCounter}. 
These functions take as input a rationalized phase space point and return numerical results where the one-loop outputs are expanded to order $\epsilon^2$.
More precisely, they return the interference of the tree-level amplitude with itself, $\mathcal{N}\operatorname{Re}\left[\overline{\sum}(\mathcal{A}^{(0)})^{\dagger}\mathcal{A}^{(0)}\right]$, with the bare one-loop amplitude, 
$2\mathcal{N}\operatorname{Re}\left[\overline{\sum}(\mathcal{A}^{(0)})^{\dagger}\mathcal{A}^{(1)}\right]$, 
and with the counter-term amplitude,
$2\mathcal{N}\operatorname{Re}\left[\overline{\sum}(\mathcal{A}^{(0)})^{\dagger}\mathcal{A}_{\rm ct.}\right]$.
The overall normalization is given by $\mathcal{N}=4\pi \alpha_s^3 y_t^2$ and $\overline{\sum}$
refers to sum and average over color and helicity states.

Furthermore, the package provides the function \texttt{NHelicityFormFactors} which returns the color-decomposed
helicity form factors, see Eq.~\eqref{eq:helicity_formfactors_even_odd} and \eqref{eq:linear_comb_hel_form_factors}.
For a detailed description of the interface, we refer the interested reader to the git repository.

For a benchmark evaluation, we fix $m_t = 175$ GeV, $y_t = 0.6914$, $\alpha_s = 0.118$,
the regularization scale $\mu=m_t$\footnote{The value of $\mu$ is kept fixed to $m_t$.} and we choose the kinematic point
\begin{align}
\begin{array}{lll}
    s_{12} =  1000000 , & \quad s_{13} =  -\dfrac{15393705013}{47152}, & \quad s_{14} = -\dfrac{39849685741}{932940}, \\[10pt]
    s_{23} = -\dfrac{21485226445}{77264},& \quad s_{24} = -\dfrac{48342263815}{112029},& \quad s_{34} =  \dfrac{83218910153}{383674}.
\end{array}
\end{align}
The tree-loop interference evaluates to 
\begin{align}
    2\mathcal{N}\operatorname{Re}&\left[\overline{\sum}(\mathcal{A}^{(0)})^{\dagger}\mathcal{A}_{\rm r}^{(1)}\right] =
    2\mathcal{N}\operatorname{Re}\left[\overline{\sum}(\mathcal{A}^{(0)})^{\dagger}\mathcal{A}^{(1)}\right] +
    2\mathcal{N}\operatorname{Re}\left[\overline{\sum}(\mathcal{A}^{(0)})^{\dagger}\mathcal{A}_{\rm ct}^{(1)}\right] =
    \notag\\
    & = \left(- \frac{0.75348873}{\epsilon^2} + \frac{1.3691456}{\epsilon} + 0.82613668 - 4.9282871 \epsilon + 1.5817369 \epsilon^2\right)\times10^{-7}.
\end{align}

The evaluation of the squared matrix element up to $\mathcal{O}(\epsilon^2)$ takes a few minutes, depending on the machine and the phase space point, for a user specified precision of $8$ digits.
We point out that the evaluation of the complete set of helicity form factors takes more time than the unpolarized
matrix element, simply because the numerical fit procedure is applied to each individual form factor instead of a single
tree-loop interference.
Approximately, $70\%$ of the total evaluation time is spent on the integration and $30\%$ is spent on the evaluation of the unreduced scattering amplitude. We expect that an implementation in \texttt{C++} with optimized pipelines could significantly improve the evaluation of the unreduced amplitude. This, however, is not straightforward as it requires a sophisticated treatment of the numerical precision. Note that from Eq.~\eqref{eq:precision_eps_fit} the numerical precision is not uniform. In other words, $8$ digits precision corresponds to the $\epsilon^2$ contribution, while the lower terms are substantially more accurate.

We performed various cross-checks of our results. First, we verified that the analytic poles of the scattering amplitude agree with the IR prediction in Eq.~\eqref{eq:Catani_Formula}. 
Moreover, we compared our numerical results up to order $\epsilon^0$ against \texttt{OpenLoops2}~\cite{Buccioni:2019sur} over a wide range of phase space points.
Furthermore, in order to verify the numerical precision of our implementation, 
we tested our numerical code in several potentially critical points characterized by a 
nearly vanishing Gram determinant built out of 2, 3 and 4 momenta. 
For all these points we found excellent agreement up to $\mathcal{O}(\epsilon^0)$ 
with a quadruple precision evaluation in \texttt{OpenLoops2}.
\section{Conclusions}

In this paper we address the calculation of the higher order terms in the $\epsilon$ expansion 
of the one-loop scattering amplitudes for  the production of a Higgs boson 
in association with a $t\bar{t}$ pair in gluon fusion.
In particular, we go beyond previous results published in~\cite{Chen:2022nxt}, by computing
fully polarized amplitudes one order higher in $\epsilon$. Our computation is 
based on a general decomposition in terms of tensor structures in the 't Hooft-Veltman scheme.
Moreover, we provide a flexible implementation of our amplitudes in the \texttt{Mathematica} package \texttt{TTH}.

The finite piece of one-loop scattering amplitudes are nowadays routinely calculated via automated providers 
in full generality. 
In this paper we started investigating the question of how an analytic approach
can cope with the complexity inherent in a five-point amplitude with massive internal and external particles. 
In this regard, 
the higher $\epsilon$ orders  serve as a proxy of the two-loop computation, as 
many salient features addressed in this article are expected to be relevant there as well. 
As an example, the  specific choice of a basis of Lorentz tensors, which makes the cancellation of 
unphysical Gram determinant denominators 
 $1/\Delta$ manifest, is a first important result of our analysis.

Although analytic results often offer great advantages in terms of performance and numerical reliability,
we argued that for such a multileg multiscale amplitude, fully symbolic results can easily become unmanageable,
even fully exploiting current technology. One of the main roadblock we have identified is the difficulty
in performing a full partial fraction decomposition of the ensuing rational functions, due to the complexity of the corresponding Groebner basis calculation.
From preliminary studies, we expect that a full partial fraction decomposition has indeed 
the potential to guarantee impressive improvements on the size of the analytic expressions.
This also suggests that it will be particularly interesting to investigate alternative methods
to reconstruct rational functions directly in partial fractioned form. 
We stress that also in this case, full control on the required Groebner basis is required.

To bypass these issues, we have developed a hybrid approach based on the use of analytic integration-by-parts identities,
concatenated with
a customized version of Auxiliary Mass Flow algorithm for the numerical integration of the master integrals. 
Given the expected drastic increase in complexity at two loops, we envision that a similar semi-numerical approach could reveal
 beneficial in this case as well.

%-----------------------------------------------------------------------------
%-----------------------------------------------------------------------------
\section*{Acknowledgments}
%-----------------------------------------------------------------------------
%-----------------------------------------------------------------------------
We thank Fabrizio Caola, Cesare Carlo Mella, Nicolas Müller, 
Dennis Ossipov, Tiziano Peraro, and Nikolaos Syrrakos for useful discussions on various aspects
of the calculation. \\
This research was partly supported by the Excellence
Cluster ORIGINS funded by the Deutsche Forschungsgemeinschaft (DFG, German Research
Foundation) under Germany's Excellence Strategy - EXC-2094 - 390783311,
and by the European Research Council (ERC) under the European Union’s research
and innovation programme grant agreements ERC Starting Grant 949279 HighPHun and 
ERC Starting Grant 804394 hipQCD. The research of XL was also supported by the UK Science and Technology Facilities Council (STFC) under grant ST/T000864/1.

%============================================
%	BIBLIOGRAPHY
%============================================
\bibliographystyle{JHEP}
\bibliography{biblio}

\providecommand{\href}[2]{#2}\begingroup\raggedright\begin{thebibliography}{10}

\bibitem{ATLAS:2012yve}
{\scshape ATLAS} collaboration, \emph{{Observation of a new particle in the
  search for the Standard Model Higgs boson with the ATLAS detector at the
  LHC}}, \href{https://doi.org/10.1016/j.physletb.2012.08.020}{\emph{Phys.
  Lett. B} {\bfseries 716} (2012) 1}
  [\href{https://arxiv.org/abs/1207.7214}{{\ttfamily 1207.7214}}].

\bibitem{CMS:2012qbp}
{\scshape CMS} collaboration, \emph{{Observation of a New Boson at a Mass of
  125 GeV with the CMS Experiment at the LHC}},
  \href{https://doi.org/10.1016/j.physletb.2012.08.021}{\emph{Phys. Lett. B}
  {\bfseries 716} (2012) 30} [\href{https://arxiv.org/abs/1207.7235}{{\ttfamily
  1207.7235}}].

\bibitem{10.1143/PTPS.1.1}
H.~Yukawa, \emph{{On the Interaction of Elementary Particles. I}},
  \href{https://doi.org/10.1143/PTPS.1.1}{\emph{Progress of Theoretical Physics
  Supplement} {\bfseries 1} (1955) 1}.

\bibitem{ATLAS:2018mme}
{\scshape ATLAS} collaboration, \emph{{Observation of Higgs boson production in
  association with a top quark pair at the LHC with the ATLAS detector}},
  \href{https://doi.org/10.1016/j.physletb.2018.07.035}{\emph{Phys. Lett. B}
  {\bfseries 784} (2018) 173}
  [\href{https://arxiv.org/abs/1806.00425}{{\ttfamily 1806.00425}}].

\bibitem{CMS:2018uxb}
{\scshape CMS} collaboration, \emph{{Observation of $\mathrm{t\overline{t}}$H
  production}},
  \href{https://doi.org/10.1103/PhysRevLett.120.231801}{\emph{Phys. Rev. Lett.}
  {\bfseries 120} (2018) 231801}
  [\href{https://arxiv.org/abs/1804.02610}{{\ttfamily 1804.02610}}].

\bibitem{ATLAS:2020ior}
{\scshape ATLAS} collaboration, \emph{{$CP$ Properties of Higgs Boson
  Interactions with Top Quarks in the $t\bar{t}H$ and $tH$ Processes Using $H
  \rightarrow \gamma\gamma$ with the ATLAS Detector}},
  \href{https://doi.org/10.1103/PhysRevLett.125.061802}{\emph{Phys. Rev. Lett.}
  {\bfseries 125} (2020) 061802}
  [\href{https://arxiv.org/abs/2004.04545}{{\ttfamily 2004.04545}}].

\bibitem{CMS:2020cga}
{\scshape CMS} collaboration, \emph{{Measurements of $\mathrm{t\bar{t}}H$
  Production and the CP Structure of the Yukawa Interaction between the Higgs
  Boson and Top Quark in the Diphoton Decay Channel}},
  \href{https://doi.org/10.1103/PhysRevLett.125.061801}{\emph{Phys. Rev. Lett.}
  {\bfseries 125} (2020) 061801}
  [\href{https://arxiv.org/abs/2003.10866}{{\ttfamily 2003.10866}}].

\bibitem{Cepeda:2019klc}
M.~Cepeda et~al., \emph{{Report from Working Group 2}: {Higgs Physics at the
  HL-LHC and HE-LHC}},
  \href{https://doi.org/10.23731/CYRM-2019-007.221}{\emph{CERN Yellow Rep.
  Monogr.} {\bfseries 7} (2019) 221}
  [\href{https://arxiv.org/abs/1902.00134}{{\ttfamily 1902.00134}}].

\bibitem{LHCHiggsCrossSectionWorkingGroup:2016ypw}
{\scshape LHC Higgs Cross Section Working Group} collaboration, \emph{{Handbook
  of LHC Higgs Cross Sections: 4. Deciphering the Nature of the Higgs Sector}},
   \href{https://arxiv.org/abs/1610.07922}{{\ttfamily 1610.07922}}.

\bibitem{PhysRevD.29.876}
J.~N. Ng and P.~Zakarauskas, \emph{Qcd-parton calculation of conjoined
  production of higgs bosons and heavy flavors in $p\overline{p}$ collisions},
  \href{https://doi.org/10.1103/PhysRevD.29.876}{\emph{Phys. Rev. D} {\bfseries
  29} (1984) 876}.

\bibitem{Kunszt:1984ri}
Z.~Kunszt, \emph{{Associated Production of Heavy Higgs Boson with Top Quarks}},
  \href{https://doi.org/10.1016/0550-3213(84)90553-4}{\emph{Nucl. Phys. B}
  {\bfseries 247} (1984) 339}.

\bibitem{Beenakker:2001rj}
W.~Beenakker, S.~Dittmaier, M.~Kramer, B.~Plumper, M.~Spira and P.~M. Zerwas,
  \emph{{Higgs radiation off top quarks at the Tevatron and the LHC}},
  \href{https://doi.org/10.1103/PhysRevLett.87.201805}{\emph{Phys. Rev. Lett.}
  {\bfseries 87} (2001) 201805}
  [\href{https://arxiv.org/abs/hep-ph/0107081}{{\ttfamily hep-ph/0107081}}].

\bibitem{Reina:2001sf}
L.~Reina and S.~Dawson, \emph{{Next-to-leading order results for t anti-t h
  production at the Tevatron}},
  \href{https://doi.org/10.1103/PhysRevLett.87.201804}{\emph{Phys. Rev. Lett.}
  {\bfseries 87} (2001) 201804}
  [\href{https://arxiv.org/abs/hep-ph/0107101}{{\ttfamily hep-ph/0107101}}].

\bibitem{Reina:2001bc}
L.~Reina, S.~Dawson and D.~Wackeroth, \emph{{QCD corrections to associated t
  anti-t h production at the Tevatron}},
  \href{https://doi.org/10.1103/PhysRevD.65.053017}{\emph{Phys. Rev. D}
  {\bfseries 65} (2002) 053017}
  [\href{https://arxiv.org/abs/hep-ph/0109066}{{\ttfamily hep-ph/0109066}}].

\bibitem{Beenakker:2002nc}
W.~Beenakker, S.~Dittmaier, M.~Kramer, B.~Plumper, M.~Spira and P.~M. Zerwas,
  \emph{{NLO QCD corrections to t anti-t H production in hadron collisions}},
  \href{https://doi.org/10.1016/S0550-3213(03)00044-0}{\emph{Nucl. Phys. B}
  {\bfseries 653} (2003) 151}
  [\href{https://arxiv.org/abs/hep-ph/0211352}{{\ttfamily hep-ph/0211352}}].

\bibitem{Dawson:2003zu}
S.~Dawson, C.~Jackson, L.~H. Orr, L.~Reina and D.~Wackeroth, \emph{{Associated
  Higgs production with top quarks at the large hadron collider: NLO QCD
  corrections}}, \href{https://doi.org/10.1103/PhysRevD.68.034022}{\emph{Phys.
  Rev. D} {\bfseries 68} (2003) 034022}
  [\href{https://arxiv.org/abs/hep-ph/0305087}{{\ttfamily hep-ph/0305087}}].

\bibitem{Frixione:2014qaa}
S.~Frixione, V.~Hirschi, D.~Pagani, H.~S. Shao and M.~Zaro, \emph{{Weak
  corrections to Higgs hadroproduction in association with a top-quark pair}},
  \href{https://doi.org/10.1007/JHEP09(2014)065}{\emph{JHEP} {\bfseries 09}
  (2014) 065} [\href{https://arxiv.org/abs/1407.0823}{{\ttfamily 1407.0823}}].

\bibitem{Zhang:2014gcy}
Y.~Zhang, W.-G. Ma, R.-Y. Zhang, C.~Chen and L.~Guo, \emph{{QCD NLO and EW NLO
  corrections to $t\bar{t}H$ production with top quark decays at hadron
  collider}}, \href{https://doi.org/10.1016/j.physletb.2014.09.022}{\emph{Phys.
  Lett. B} {\bfseries 738} (2014) 1}
  [\href{https://arxiv.org/abs/1407.1110}{{\ttfamily 1407.1110}}].

\bibitem{Frixione:2015zaa}
S.~Frixione, V.~Hirschi, D.~Pagani, H.~S. Shao and M.~Zaro, \emph{{Electroweak
  and QCD corrections to top-pair hadroproduction in association with heavy
  bosons}}, \href{https://doi.org/10.1007/JHEP06(2015)184}{\emph{JHEP}
  {\bfseries 06} (2015) 184}
  [\href{https://arxiv.org/abs/1504.03446}{{\ttfamily 1504.03446}}].

\bibitem{Denner:2016wet}
A.~Denner, J.-N. Lang, M.~Pellen and S.~Uccirati, \emph{{Higgs production in
  association with off-shell top-antitop pairs at NLO EW and QCD at the LHC}},
  \href{https://doi.org/10.1007/JHEP02(2017)053}{\emph{JHEP} {\bfseries 02}
  (2017) 053} [\href{https://arxiv.org/abs/1612.07138}{{\ttfamily
  1612.07138}}].

\bibitem{Catani:2021cbl}
S.~Catani, I.~Fabre, M.~Grazzini and S.~Kallweit, \emph{{${t {{\bar{t}}}H}$
  production at NNLO: the flavour off-diagonal channels}},
  \href{https://doi.org/10.1140/epjc/s10052-021-09247-w}{\emph{Eur. Phys. J. C}
  {\bfseries 81} (2021) 491}
  [\href{https://arxiv.org/abs/2102.03256}{{\ttfamily 2102.03256}}].

\bibitem{Catani:2022mfv}
S.~Catani, S.~Devoto, M.~Grazzini, S.~Kallweit, J.~Mazzitelli and C.~Savoini,
  \emph{{Higgs Boson Production in Association with a Top-Antitop Quark Pair in
  Next-to-Next-to-Leading Order QCD}},
  \href{https://doi.org/10.1103/PhysRevLett.130.111902}{\emph{Phys. Rev. Lett.}
  {\bfseries 130} (2023) 111902}
  [\href{https://arxiv.org/abs/2210.07846}{{\ttfamily 2210.07846}}].

\bibitem{Agarwal:2021vdh}
B.~Agarwal, F.~Buccioni, A.~von Manteuffel and L.~Tancredi, \emph{{Two-Loop
  Helicity Amplitudes for Diphoton Plus Jet Production in Full Color}},
  \href{https://doi.org/10.1103/PhysRevLett.127.262001}{\emph{Phys. Rev. Lett.}
  {\bfseries 127} (2021) 262001}
  [\href{https://arxiv.org/abs/2105.04585}{{\ttfamily 2105.04585}}].

\bibitem{Badger:2021imn}
S.~Badger, C.~Br\o{}nnum-Hansen, D.~Chicherin, T.~Gehrmann, H.~B. Hartanto,
  J.~Henn et~al., \emph{{Virtual QCD corrections to gluon-initiated diphoton
  plus jet production at hadron colliders}},
  \href{https://doi.org/10.1007/JHEP11(2021)083}{\emph{JHEP} {\bfseries 11}
  (2021) 083} [\href{https://arxiv.org/abs/2106.08664}{{\ttfamily
  2106.08664}}].

\bibitem{Abreu:2023bdp}
S.~Abreu, G.~De~Laurentis, H.~Ita, M.~Klinkert, B.~Page and V.~Sotnikov,
  \emph{{Two-loop QCD corrections for three-photon production at hadron
  colliders}},
  \href{https://doi.org/10.21468/SciPostPhys.15.4.157}{\emph{SciPost Phys.}
  {\bfseries 15} (2023) 157}
  [\href{https://arxiv.org/abs/2305.17056}{{\ttfamily 2305.17056}}].

\bibitem{Badger:2023mgf}
S.~Badger, M.~Czakon, H.~B. Hartanto, R.~Moodie, T.~Peraro, R.~Poncelet et~al.,
  \emph{{Isolated photon production in association with a jet pair through
  next-to-next-to-leading order in QCD}},
  \href{https://doi.org/10.1007/JHEP10(2023)071}{\emph{JHEP} {\bfseries 10}
  (2023) 071} [\href{https://arxiv.org/abs/2304.06682}{{\ttfamily
  2304.06682}}].

\bibitem{Agarwal:2023suw}
B.~Agarwal, F.~Buccioni, F.~Devoto, G.~Gambuti, A.~von Manteuffel and
  L.~Tancredi, \emph{{Five-Parton Scattering in QCD at Two Loops}},
  \href{https://arxiv.org/abs/2311.09870}{{\ttfamily 2311.09870}}.

\bibitem{DeLaurentis:2023nss}
G.~De~Laurentis, H.~Ita, M.~Klinkert and V.~Sotnikov, \emph{{Double-Virtual
  NNLO QCD Corrections for Five-Parton Scattering: The Gluon Channel}},
  \href{https://arxiv.org/abs/2311.10086}{{\ttfamily 2311.10086}}.

\bibitem{DeLaurentis:2023izi}
G.~De~Laurentis, H.~Ita and V.~Sotnikov, \emph{{Double-Virtual NNLO QCD
  Corrections for Five-Parton Scattering: The Quark Channels}},
  \href{https://arxiv.org/abs/2311.18752}{{\ttfamily 2311.18752}}.

\bibitem{Badger:2021nhg}
S.~Badger, H.~B. Hartanto and S.~Zoia, \emph{{Two-Loop QCD Corrections to
  Wbb\textasciimacron{} Production at Hadron Colliders}},
  \href{https://doi.org/10.1103/PhysRevLett.127.012001}{\emph{Phys. Rev. Lett.}
  {\bfseries 127} (2021) 012001}
  [\href{https://arxiv.org/abs/2102.02516}{{\ttfamily 2102.02516}}].

\bibitem{Abreu:2021asb}
S.~Abreu, F.~Febres~Cordero, H.~Ita, M.~Klinkert, B.~Page and V.~Sotnikov,
  \emph{{Leading-color two-loop amplitudes for four partons and a W boson in
  QCD}}, \href{https://doi.org/10.1007/JHEP04(2022)042}{\emph{JHEP} {\bfseries
  04} (2022) 042} [\href{https://arxiv.org/abs/2110.07541}{{\ttfamily
  2110.07541}}].

\bibitem{Badger:2021ega}
S.~Badger, H.~B. Hartanto, J.~Kry\'s and S.~Zoia, \emph{{Two-loop
  leading-colour QCD helicity amplitudes for Higgs boson production in
  association with a bottom-quark pair at the LHC}},
  \href{https://doi.org/10.1007/JHEP11(2021)012}{\emph{JHEP} {\bfseries 11}
  (2021) 012} [\href{https://arxiv.org/abs/2107.14733}{{\ttfamily
  2107.14733}}].

\bibitem{Badger:2022ncb}
S.~Badger, H.~B. Hartanto, J.~Kry\'s and S.~Zoia, \emph{{Two-loop leading
  colour helicity amplitudes for $W^{\pm}\gamma + j$ production at the LHC}},
  \href{https://doi.org/10.1007/JHEP05(2022)035}{\emph{JHEP} {\bfseries 05}
  (2022) 035} [\href{https://arxiv.org/abs/2201.04075}{{\ttfamily
  2201.04075}}].

\bibitem{Abreu:2023rco}
S.~Abreu, D.~Chicherin, H.~Ita, B.~Page, V.~Sotnikov, W.~Tschernow et~al.,
  \emph{{All Two-Loop Feynman Integrals for Five-Point One-Mass Scattering}},
  \href{https://arxiv.org/abs/2306.15431}{{\ttfamily 2306.15431}}.

\bibitem{Badger:2022mrb}
S.~Badger, M.~Becchetti, E.~Chaubey, R.~Marzucca and F.~Sarandrea,
  \emph{{One-loop QCD helicity amplitudes for pp \textrightarrow{} $
  t\overline{t}j $ to O(\ensuremath{\varepsilon}$^{2}$)}},
  \href{https://doi.org/10.1007/JHEP06(2022)066}{\emph{JHEP} {\bfseries 06}
  (2022) 066} [\href{https://arxiv.org/abs/2201.12188}{{\ttfamily
  2201.12188}}].

\bibitem{Badger:2022hno}
S.~Badger, M.~Becchetti, E.~Chaubey and R.~Marzucca, \emph{{Two-loop master
  integrals for a planar topology contributing to pp \textrightarrow{}$
  t\overline{t}j $}},
  \href{https://doi.org/10.1007/JHEP01(2023)156}{\emph{JHEP} {\bfseries 01}
  (2023) 156} [\href{https://arxiv.org/abs/2210.17477}{{\ttfamily
  2210.17477}}].

\bibitem{FebresCordero:2023gjh}
F.~Febres~Cordero, G.~Figueiredo, M.~Kraus, B.~Page and L.~Reina,
  \emph{{Two-Loop Master Integrals for Leading-Color $pp\to t\bar{t}H$
  Amplitudes with a Light-Quark Loop}},
  \href{https://arxiv.org/abs/2312.08131}{{\ttfamily 2312.08131}}.

\bibitem{Chen:2022nxt}
J.~Chen, C.~Ma, G.~Wang, L.~L. Yang and X.~Ye, \emph{{Two-loop infrared
  singularities in the production of a Higgs boson associated with a top-quark
  pair}}, \href{https://doi.org/10.1007/JHEP04(2022)025}{\emph{JHEP} {\bfseries
  04} (2022) 025} [\href{https://arxiv.org/abs/2202.02913}{{\ttfamily
  2202.02913}}].

\bibitem{Henn:2013pwa}
J.~M. Henn, \emph{{Multiloop integrals in dimensional regularization made
  simple}}, \href{https://doi.org/10.1103/PhysRevLett.110.251601}{\emph{Phys.
  Rev. Lett.} {\bfseries 110} (2013) 251601}
  [\href{https://arxiv.org/abs/1304.1806}{{\ttfamily 1304.1806}}].

\bibitem{Tkachov:1981wb}
F.~V. Tkachov, \emph{{A theorem on analytical calculability of 4-loop
  renormalization group functions}},
  \href{https://doi.org/10.1016/0370-2693(81)90288-4}{\emph{Phys. Lett. B}
  {\bfseries 100} (1981) 65}.

\bibitem{Chetyrkin:1981qh}
K.~G. Chetyrkin and F.~V. Tkachov, \emph{{Integration by parts: The algorithm
  to calculate $\beta$-functions in 4 loops}},
  \href{https://doi.org/10.1016/0550-3213(81)90199-1}{\emph{Nucl. Phys. B}
  {\bfseries 192} (1981) 159}.

\bibitem{RevModPhys.36.595}
N.~Byers and C.~N. Yang, \emph{Physical regions in invariant variables for $n$
  particles and the phase-space volume element},
  \href{https://doi.org/10.1103/RevModPhys.36.595}{\emph{Rev. Mod. Phys.}
  {\bfseries 36} (1964) 595}.

\bibitem{tHooft:1972tcz}
G.~'t~Hooft and M.~J.~G. Veltman, \emph{{Regularization and Renormalization of
  Gauge Fields}},
  \href{https://doi.org/10.1016/0550-3213(72)90279-9}{\emph{Nucl. Phys. B}
  {\bfseries 44} (1972) 189}.

\bibitem{Peraro:2019cjj}
T.~Peraro and L.~Tancredi, \emph{{Physical projectors for multi-leg helicity
  amplitudes}}, \href{https://doi.org/10.1007/JHEP07(2019)114}{\emph{JHEP}
  {\bfseries 07} (2019) 114}
  [\href{https://arxiv.org/abs/1906.03298}{{\ttfamily 1906.03298}}].

\bibitem{Peraro:2020sfm}
T.~Peraro and L.~Tancredi, \emph{{Tensor decomposition for bosonic and
  fermionic scattering amplitudes}},
  \href{https://doi.org/10.1103/PhysRevD.103.054042}{\emph{Phys. Rev. D}
  {\bfseries 103} (2021) 054042}
  [\href{https://arxiv.org/abs/2012.00820}{{\ttfamily 2012.00820}}].

\bibitem{Dixon:1996wi}
L.~J. Dixon, \emph{{Calculating scattering amplitudes efficiently}},  in
  \emph{{Theoretical Advanced Study Institute in Elementary Particle Physics
  (TASI 95): QCD and Beyond}}, pp.~539--584, 1, 1996,
  \href{https://arxiv.org/abs/hep-ph/9601359}{{\ttfamily hep-ph/9601359}}.

\bibitem{Arkani-Hamed:2017jhn}
N.~Arkani-Hamed, T.-C. Huang and Y.-t. Huang, \emph{{Scattering amplitudes for
  all masses and spins}},
  \href{https://doi.org/10.1007/JHEP11(2021)070}{\emph{JHEP} {\bfseries 11}
  (2021) 070} [\href{https://arxiv.org/abs/1709.04891}{{\ttfamily
  1709.04891}}].

\bibitem{Badger:2021owl}
S.~Badger, E.~Chaubey, H.~B. Hartanto and R.~Marzucca, \emph{{Two-loop leading
  colour QCD helicity amplitudes for top quark pair production in the gluon
  fusion channel}}, \href{https://doi.org/10.1007/JHEP06(2021)163}{\emph{JHEP}
  {\bfseries 06} (2021) 163}
  [\href{https://arxiv.org/abs/2102.13450}{{\ttfamily 2102.13450}}].

\bibitem{Denner:2019vbn}
A.~Denner and S.~Dittmaier, \emph{{Electroweak Radiative Corrections for
  Collider Physics}},
  \href{https://doi.org/10.1016/j.physrep.2020.04.001}{\emph{Phys. Rept.}
  {\bfseries 864} (2020) 1} [\href{https://arxiv.org/abs/1912.06823}{{\ttfamily
  1912.06823}}].

\bibitem{Melnikov:2000zc}
K.~Melnikov and T.~van Ritbergen, \emph{{The Three loop on-shell
  renormalization of QCD and QED}},
  \href{https://doi.org/10.1016/S0550-3213(00)00526-5}{\emph{Nucl. Phys. B}
  {\bfseries 591} (2000) 515}
  [\href{https://arxiv.org/abs/hep-ph/0005131}{{\ttfamily hep-ph/0005131}}].

\bibitem{Catani:1998bh}
S.~Catani, \emph{{The Singular behavior of QCD amplitudes at two loop order}},
  \href{https://doi.org/10.1016/S0370-2693(98)00332-3}{\emph{Phys. Lett. B}
  {\bfseries 427} (1998) 161}
  [\href{https://arxiv.org/abs/hep-ph/9802439}{{\ttfamily hep-ph/9802439}}].

\bibitem{Catani:2002hc}
S.~Catani, S.~Dittmaier, M.~H. Seymour and Z.~Trocsanyi, \emph{{The Dipole
  formalism for next-to-leading order QCD calculations with massive partons}},
  \href{https://doi.org/10.1016/S0550-3213(02)00098-6}{\emph{Nucl. Phys. B}
  {\bfseries 627} (2002) 189}
  [\href{https://arxiv.org/abs/hep-ph/0201036}{{\ttfamily hep-ph/0201036}}].

\bibitem{Becher:2009cu}
T.~Becher and M.~Neubert, \emph{{Infrared singularities of scattering
  amplitudes in perturbative QCD}},
  \href{https://doi.org/10.1103/PhysRevLett.102.162001}{\emph{Phys. Rev. Lett.}
  {\bfseries 102} (2009) 162001}
  [\href{https://arxiv.org/abs/0901.0722}{{\ttfamily 0901.0722}}].

\bibitem{Becher:2009kw}
T.~Becher and M.~Neubert, \emph{{Infrared singularities of QCD amplitudes with
  massive partons}},
  \href{https://doi.org/10.1103/PhysRevD.79.125004}{\emph{Phys. Rev. D}
  {\bfseries 79} (2009) 125004}
  [\href{https://arxiv.org/abs/0904.1021}{{\ttfamily 0904.1021}}].

\bibitem{Nogueira:1991ex}
P.~Nogueira, \emph{{Automatic Feynman Graph Generation}},
  \href{https://doi.org/10.1006/jcph.1993.1074}{\emph{J. Comput. Phys.}
  {\bfseries 105} (1993) 279}.

\bibitem{Ruijl:2017dtg}
B.~Ruijl, T.~Ueda and J.~Vermaseren, \emph{{FORM version 4.2}},
  \href{https://arxiv.org/abs/1707.06453}{{\ttfamily 1707.06453}}.

\bibitem{Studerus:2009ye}
C.~Studerus, \emph{{Reduze-Feynman Integral Reduction in C++}},
  \href{https://doi.org/10.1016/j.cpc.2010.03.012}{\emph{Comput. Phys. Commun.}
  {\bfseries 181} (2010) 1293}
  [\href{https://arxiv.org/abs/0912.2546}{{\ttfamily 0912.2546}}].

\bibitem{Maierhofer:2017gsa}
P.~Maierh\"ofer, J.~Usovitsch and P.~Uwer, \emph{{Kira\textemdash{}A Feynman
  integral reduction program}},
  \href{https://doi.org/10.1016/j.cpc.2018.04.012}{\emph{Comput. Phys. Commun.}
  {\bfseries 230} (2018) 99}
  [\href{https://arxiv.org/abs/1705.05610}{{\ttfamily 1705.05610}}].

\bibitem{Klappert:2020nbg}
J.~Klappert, F.~Lange, P.~Maierh\"ofer and J.~Usovitsch, \emph{{Integral
  reduction with Kira 2.0 and finite field methods}},
  \href{https://doi.org/10.1016/j.cpc.2021.108024}{\emph{Comput. Phys. Commun.}
  {\bfseries 266} (2021) 108024}
  [\href{https://arxiv.org/abs/2008.06494}{{\ttfamily 2008.06494}}].

\bibitem{Liu:2022chg}
X.~Liu and Y.-Q. Ma, \emph{{AMFlow: A Mathematica package for Feynman integrals
  computation via auxiliary mass flow}},
  \href{https://doi.org/10.1016/j.cpc.2022.108565}{\emph{Comput. Phys. Commun.}
  {\bfseries 283} (2023) 108565}
  [\href{https://arxiv.org/abs/2201.11669}{{\ttfamily 2201.11669}}].

\bibitem{Ferguson:PSLQ}
H.~Ferguson and D.~Bailey, \emph{A polynomial time, numerically stable integer
  relation algorithm}, {\emph{RNR Technical Report RNR-91-032} (1992) }.

\bibitem{Bogner:2010kv}
C.~Bogner and S.~Weinzierl, \emph{{Feynman graph polynomials}},
  \href{https://doi.org/10.1142/S0217751X10049438}{\emph{Int. J. Mod. Phys. A}
  {\bfseries 25} (2010) 2585}
  [\href{https://arxiv.org/abs/1002.3458}{{\ttfamily 1002.3458}}].

\bibitem{Adams:2018bsn}
L.~Adams, E.~Chaubey and S.~Weinzierl, \emph{{Planar Double Box Integral for
  Top Pair Production with a Closed Top Loop to all orders in the Dimensional
  Regularization Parameter}},
  \href{https://doi.org/10.1103/PhysRevLett.121.142001}{\emph{Phys. Rev. Lett.}
  {\bfseries 121} (2018) 142001}
  [\href{https://arxiv.org/abs/1804.11144}{{\ttfamily 1804.11144}}].

\bibitem{lewis_fermat_2008}
R.~H. Lewis, \emph{{Fermat computer algebra syetem}}, {\emph{Mathematics
  Department, Fordham University} (2008) }
  [\href{https://arxiv.org/abs/http://home.bway.net/lewis/}{{\ttfamily
  http://home.bway.net/lewis/}}].

\bibitem{vonManteuffel:2014ixa}
A.~von Manteuffel and R.~M. Schabinger, \emph{{A novel approach to integration
  by parts reduction}},
  \href{https://doi.org/10.1016/j.physletb.2015.03.029}{\emph{Phys. Lett. B}
  {\bfseries 744} (2015) 101}
  [\href{https://arxiv.org/abs/1406.4513}{{\ttfamily 1406.4513}}].

\bibitem{Peraro:2019svx}
T.~Peraro, \emph{{FiniteFlow: multivariate functional reconstruction using
  finite fields and dataflow graphs}},
  \href{https://doi.org/10.1007/JHEP07(2019)031}{\emph{JHEP} {\bfseries 07}
  (2019) 031} [\href{https://arxiv.org/abs/1905.08019}{{\ttfamily
  1905.08019}}].

\bibitem{Heller:2021qkz}
M.~Heller and A.~von Manteuffel, \emph{{MultivariateApart: Generalized partial
  fractions}}, \href{https://doi.org/10.1016/j.cpc.2021.108174}{\emph{Comput.
  Phys. Commun.} {\bfseries 271} (2022) 108174}
  [\href{https://arxiv.org/abs/2101.08283}{{\ttfamily 2101.08283}}].

\bibitem{DGPS}
W.~Decker, G.-M. Greuel, G.~Pfister and H.~Sch\"onemann, ``{\sc Singular}
  {4-2-0} --- {A} computer algebra system for polynomial computations.''
  \url{http://www.singular.uni-kl.de}, 2020.

\bibitem{Magerya:2022hvj}
V.~Magerya, \emph{{Rational Tracer: a Tool for Faster Rational Function
  Reconstruction}},  \href{https://arxiv.org/abs/2211.03572}{{\ttfamily
  2211.03572}}.

\bibitem{ZIPPEL1990375}
R.~Zippel, \emph{Interpolating polynomials from their values},
  \href{https://doi.org/https://doi.org/10.1016/S0747-7171(08)80018-1}{\emph{Journal
  of Symbolic Computation} {\bfseries 9} (1990) 375}.

\bibitem{Chawdhry:2023yyx}
H.~A. Chawdhry, \emph{{p-adic reconstruction of rational functions in
  multi-loop amplitudes}},  \href{https://arxiv.org/abs/2312.03672}{{\ttfamily
  2312.03672}}.

\bibitem{Kreer:2021sdt}
P.~A. Kreer and S.~Weinzierl, \emph{{The H-graph with equal masses in terms of
  multiple polylogarithms}},
  \href{https://doi.org/10.1016/j.physletb.2021.136405}{\emph{Phys. Lett. B}
  {\bfseries 819} (2021) 136405}
  [\href{https://arxiv.org/abs/2104.07488}{{\ttfamily 2104.07488}}].

\bibitem{Liu:2017jxz}
X.~Liu, Y.-Q. Ma and C.-Y. Wang, \emph{{A Systematic and Efficient Method to
  Compute Multi-loop Master Integrals}},
  \href{https://doi.org/10.1016/j.physletb.2018.02.026}{\emph{Phys. Lett. B}
  {\bfseries 779} (2018) 353}
  [\href{https://arxiv.org/abs/1711.09572}{{\ttfamily 1711.09572}}].

\bibitem{Lee:2013mka}
R.~N. Lee, \emph{{LiteRed 1.4: a powerful tool for reduction of multiloop
  integrals}}, \href{https://doi.org/10.1088/1742-6596/523/1/012059}{\emph{J.
  Phys. Conf. Ser.} {\bfseries 523} (2014) 012059}
  [\href{https://arxiv.org/abs/1310.1145}{{\ttfamily 1310.1145}}].

\bibitem{Smirnov:2019qkx}
A.~V. Smirnov and F.~S. Chuharev, \emph{{FIRE6: Feynman Integral REduction with
  Modular Arithmetic}},
  \href{https://doi.org/10.1016/j.cpc.2019.106877}{\emph{Comput. Phys. Commun.}
  {\bfseries 247} (2020) 106877}
  [\href{https://arxiv.org/abs/1901.07808}{{\ttfamily 1901.07808}}].

\bibitem{Duplancic:2003tv}
G.~Duplancic and B.~Nizic, \emph{{Reduction method for dimensionally regulated
  one loop N point Feynman integrals}},
  \href{https://doi.org/10.1140/epjc/s2004-01723-7}{\emph{Eur. Phys. J. C}
  {\bfseries 35} (2004) 105}
  [\href{https://arxiv.org/abs/hep-ph/0303184}{{\ttfamily hep-ph/0303184}}].

\bibitem{Buccioni:2019sur}
F.~Buccioni, J.-N. Lang, J.~M. Lindert, P.~Maierh\"ofer, S.~Pozzorini, H.~Zhang
  et~al., \emph{{OpenLoops 2}},
  \href{https://doi.org/10.1140/epjc/s10052-019-7306-2}{\emph{Eur. Phys. J. C}
  {\bfseries 79} (2019) 866}
  [\href{https://arxiv.org/abs/1907.13071}{{\ttfamily 1907.13071}}].

\end{thebibliography}\endgroup

\end{document}